\documentclass[preprint,12pt,authoryear]{elsarticle}%

\setcitestyle{numbers}
\usepackage{adjustbox}
\usepackage{subfigure}
\usepackage{amsmath}
\usepackage{amssymb}
\usepackage{amstext}
\usepackage{array,booktabs}
\usepackage{array}
\usepackage{blindtext}
\usepackage{bm}
\usepackage{color}
\usepackage{comment}
\usepackage{dcolumn}
\usepackage{enumitem}
\usepackage{fancyhdr}
\usepackage{float}
\usepackage[dvipsnames]{xcolor}
\usepackage[margin=1in]{geometry}
\usepackage{graphics}
\usepackage{hyperref}
\usepackage{mathrsfs}
\usepackage{mathtools}
\usepackage{multirow}
\usepackage{pgfplots}
\usepackage{scalerel}
\usepackage{stackengine,wasysym}

\usepackage{psfrag}
\usepackage{scalerel}
\usepackage{siunitx}
\usepackage{stackengine,wasysym}
\usepackage{subcaption}
\usepackage{threeparttable}
\usepackage{tikz}
\usepackage{titlesec}
\usepackage{upgreek}
\usepgfplotslibrary{fillbetween}
\usepackage[toc,page]{appendix}

\definecolor{myOrange}{HTML}{E69F00}
\definecolor{skyBlue}{HTML}{56b4e9}
\definecolor{blueGreen}{HTML}{009e73}
\definecolor{myBlue}{HTML}{0072b2}
\definecolor{vermillion}{HTML}{d55e00}
\definecolor{paleViolet}{HTML}{cc79a7}

\providecommand{\nomname}{}  

\newlength\nomenwidth
\newlength\savitemsep

\NewDocumentCommand\entry{m m m}{% simplified greatly 2024/01/18
             \item[#1\hfill]\begin{tabular}[t]{p{0.7\linewidth} l}#2 & #3\end{tabular}%
             \@itempenalty=-\@lowpenalty
}

\NewDocumentCommand\EntryHeading{m}{%
	\itemsep3\p@ plus 1\p@ minus 1\p@
    \goodbreak\item[\itshape#1\hfill]\mbox{}%
    \setlength{\itemsep}{\savitemsep}\@itempenalty=1000
}

\NewDocumentEnvironment{nomenclature}{O{2.5em} O{\nomname}}{%
        \setlength{\columnsep}{2em} 
        \setlength{\nomenwidth}{#1}
        \section*{#2}
        \raggedright
        \begin{list}{}{%
             \setlength{\itemsep}{0pt}%
             \setlength{\parsep}{\itemsep}%
             \setlength{\labelsep}{1em}%
             \setlength{\labelwidth}{\nomenwidth}%
             \setlength{\leftmargin}{\labelwidth}%
             \addtolength{\leftmargin}{\labelsep}%
			 \setlength{\savitemsep}{\itemsep}%
        }%
}{\end{list}\ignorespacesafterend} 
\pgfplotsset{
        layers/my layer set/.define layer set={
            background,
            main,
            foreground
        }{
           
        },
        set layers=my layer set,    }

\journal{International Journal of Multiphase Flow}
\numberwithin{equation}{section}
\newcolumntype{P}[1]{>{\centering\arraybackslash}p{#1}}
\begin{document}
\begin{frontmatter}

\title{A filter-dependent granular temperature model from large-scale CFD-DEM data}

\author[label1]{L. Rosenberg}
\author[label2]{W.Fullmer}
\author[label1]{S. Beetham\footnote{Cooresponding author: sbeetham@oakland.edu}}
\address[label1]{Oakland University, Department of Mechanical Engineering}
\address[label2]{National Energy Technology Laboratory}

\begin{abstract}
\label{abstract}
The computational study of strongly-coupled, gas-solid flows at scales relevant to most environmental and engineering applications requires the use of `coarse-grained' methodologies such as the two-fluid model, particle-in-cell approach or the multiphase Reynolds Averaged Navier--Stokes equations. While these strategies enable computations at desirable length- and time-scales, they rely heavily on models to capture important flow physics that occur at scales smaller than the mesh. To date, the models that do exist are based on a limited set of flow conditions, such as very dilute particle phase. To this end, we leverage the most large-scale repository of CFD-DEM data to date to develop a filter-size dependent model for granular temperature--a key quantity for accurately predicting gas-solid flows. This filter-size dependence then translates directly to grid-size dependence in the context of coarse-grained approaches.  In addition, we leverage the CFD-DEM dataset to propose an improved model for the mean variance in particle volume fraction, a quantity that characterizes the degree of clustering in the flow. 

%The development of an aggregate model that captures the impact of varied Gaussian filtering of CFD-DEM data on granular temperature has not yet been investigated on a large scale. The impact on sub-grid turbulent energy is of particular interest in schemes such as CFD-DEM, as the grid must be of large enough granularity to accommodate the calculation of interphase exchange terms, therefore the subcomponents of particle granular energy are left wholly at the mercy of models. Herein we explore multiple models that take the particle Reynold's number, nondimensional filter width, and other select Reynolds averaged quantities to represent volume fraction fluctuations, a partitioning for particle granular energy subcomponents, and a granular temperature based Reynolds number.
\end{abstract}

\begin{keyword}
Gas-solid flows \sep particle-laden \sep data-driven modeling \sep  cluster induced turbulence
\end{keyword}

\end{frontmatter}

\section{Introduction}
\label{sec:Introduction}
% 1. Gas-solid flows are pervasive/ubiquitous in natural and industrial flows of importance and these problems often are large (domain sizes are large, numbers of particles are large).
Gravity-driven, gas-solid flows are ubiquitous across both industrial and natural processes. For many of these flows, improving understanding of underlying flow physics and advancing our ability to tractably and accurately predict them is often either industrially relevant or socially critical. For instance, pyroclastic density currents are the gravity-driven flow of ash particles in hot gas and represent the most hazardous volcanic process~\cite{foster2025settling}. Gas-solid flows are also prevalent in a range of chemical engineering applications, including the circulating fluidized reactors used to upgrade various feedstock into fuel\cite{beetham2019biomass}\cite{groll1993reaction}. 

In nearly all applications of interest, an expansive range of both length and time scales is present and the physics proceeding at the small scales have important implications on large-scale behavior. This reverse cascade is in direct opposition to the traditional energy cascade established in single-phase turbulence (K41) theory~\cite{kolmogorov1941dissipation}. To this end, when particles are fluidized by a gas, the resulting multiphase flow is almost always complex and chaotic, generating heterogeneous structures commonly known as clusters in particle-dilute flows and bubbles in particle-dense flows (\citet{agrawal2001role}; \citet{fullmer2018continuum}). These structures often cause the development of turbulence in the gas-phase, which would not have otherwise been present in the absence of a particulate phase (sometimes termed `cluster induced turbulence'~\cite{capecelatro2014investigating}). 

% 2. High fidelity computational methods exist for simulating gas-solid flows (PR-DNS, EL), but they are limited in size. Therefore, scale up is needed to predict such flows at relevant environmental/industrial scales. 3. Approaches that are tractable at scale (PIC, EE) require a heavy reliance on models of sub filter or sub grid scale physics. To date, accurate models for these approaches remain elusive or untested. 
For this class of flows, several approaches exist with varying degrees of accuracy and tractability. Particle-resolved Direct Numerical Simulation (PR-DNS)\cite{tenneti2014particle}, allows for the resolution of particle-scale flow physics and requires very little modeling (e.g., only chemical kinetics, lubrication forces and collisions are modeled). While highly accurate, this approach is extremely limited to the number of particles in the system ($\mathcal{O}(10^3)$), which in turn limits the domain size due to sub-particle scale resolution requirements, thus making PR-DNS intractable for systems of industrial interest. In order to enable the tractability of systems that are capable of exhibiting meso-scale structures such as clusters and bubbling, methods such as unresolved Euler--Lagrange (EL), commonly referred to as Computational Fluid Dynamics--Discrete Element Method (CFD-DEM), can be employed where models are introduced for particle-scale physics (e.g., drag models)\cite{tenneti2011drag}. While these approaches enable the simulation of much more expansive systems, they still fall short of tractably capturing scales of interest. Extending the reliance on models even further results in techniques such as two-fluid or Euler-Euler (EE) models or Particle-In-Cell approaches (PIC). The former treats both phases in the Eulerian sense and requires several closures for the particle-phase, particularly for drag, granular temperature and the particle stress tensor. While methods such as kinetic-theory based two-fluid models (e.g.,\citet{garzo2012}) are able to capture complex interactions between both phases and the development of heterogeneous structures, it takes a significant amount of resolution to do so~\citet{fullmer2016quantitative}, \citet{fullmer2017clustering}, making these approaches insufficient for capturing large-scale industrially-relevant problems. In the case of PIC, Lagrangian markers that exist at discrete points in space and time are used to represent parcels of potentially non-zero, non-integer numbers of particles. These parcels do not directly interact with each other, but instead communicate via a solids stress tensor that is computed based on the volume fraction\cite{ANDREWS1996379}\cite{snider2001incompressible}\cite{clarke2020mfix}.

Given this context, improved models are the key hurdle preventing coarse grained approaches from accurately resolving small-scale dynamics. Thus far, modeling has proved challenging with most efforts focused on the filtered (unresolved) drag force, e.g., \citet{igci2008filtered}, \citet{igci2011constitutive}, \citet{sarkar2016filtered}, \citet{ozel2017towards}, among many others. However, in addition to modeling the sub-grid drag force, another quantity of interest is also important for gas-solid flows--the granular temperature, $\Theta$, a measure of the unresolved granular energy~\cite{GOLDHIRSCH2008130}. 

%4. Large amounts of high-fidelity data is needed across a wide parameter space to inform more accurate models. To this end, a recent large scale EL simulation campaign was carried out (this is the ALCC data). This work leverages this extensive dataset and builds upon the existing Tang model for granular temperature. The resulting model is dependent upon filter size and represents a first step towards more accurate models that are sensitive to filter or grid size. 

In order to build a model that is more likely to be representative of a wide range of flow states and regimes, it is important to take into account highly-resolved data across each of these regions. To this end, the most expansive set of gravity-driven, gas-solid CFD-DEM data to date has been curated and used in this work to develop a filter-dependent granular temperature model. To this end, this paper is structured in the following manner: First, the concept of granular temperature and its relation to other fluctuating kinetic energy terms in the particle field is introduced in Sec.~\ref{sec:theory}. Next, the system and properties of the CFD-DEM simulations used to produce the dataset used in this work is provided in Sec.~\ref{sec:method}. Finally, the data is analyzed in Sec.~\ref{sec:results} culminating in the proposed granular temperature model, before ending with summarizing remarks in Sec.~\ref{sec:conclusions}.

\section{A brief formalization of granular temperature}\label{sec:theory}
While several extensive studies of granular temperature and its origins can be found in the literature~\cite{fullmer2017a}, here we present a brief overview to support the modeling discussions presented in this work. 

The concept of granular temperature is rooted in Einstein's 1905 postulation that macroscale dynamics of particles proportionally scales to molecular dynamics \cite{einstein1905movement}\cite{GOLDHIRSCH2008130}. Fluidized particulate (or any kind of granular gas) effectively embodies the same scaling of molecular behavior as a standard gas, however with some key nuances. Inter-particle collisions contribute to granular temperature in the same paradigm as inter-molecule collisions do in any other substance; the more collisions, the higher the rate of change of entropy versus the change in internal energy, and subsequently the higher the temperature. However in a departure from traditional molecular dynamics, inter-particle collisions are inelastic, therefore there is a requirement of a source of energy in a fluidized system to maintain the fluidization\cite{GOLDHIRSCH2008130}.

In CFD-DEM, two-phase flows are solved by evolving the fluid on a mesh in the usual Eulerian sense and evolving the particles as discrete Lagrangian bodies subject to Newton's second law. The phases are then coupled via interphase exchange terms~\cite{capecelatro2013euler}. Due to this, Lagrangian quantities, such as the particle-phase velocity, can be decomposed as 
\begin{equation}
    \bm{u}_p = \tilde{\bm{u}}_p + \delta \bm{u}_p,
\end{equation}
where $\bm{u}_p$ denotes the total, instantaneous particle velocity, $\tilde{\bm{u}}_p$ is the \emph{Eulerian} particle velocity and $\delta \bm{u}_p$ is an uncorrelated, Lagrangian residual. The procedure for generating an Eulerian representation for Lagrangian quantities involves a two-step filtering operation and is described in detail in the literature~\cite{capecelatro2013euler} as well as Sec.~\ref{sec:method}.

%In this context, granular temperature can be thought of as the \emph{unresolved} contribution to total granular energy. In order to properly develop its definition, it is important to explain how Lagrangian (particle) information is communicated to the Eulerian (fluid) grid as well as define relevant averaging procedures.  

In this work, a Reynolds average is denoted with angled brackets, $\langle \cdot \rangle$, and fluctuations from Reynolds averaged quantities are denoted with a single prime. It is notable that because the simulations carried out in this study are triply periodic and become statistically stationary in time, this average is carried out over all three spatial dimensions as well as time. Given this, the decomposition of the particle velocity, $\bm{u}_p$, is given as $\bm{u}_p = \langle \bm{u}_p \rangle + \bm{u}_p^{\prime}$ and the decomposition of the fluid velocity, $\bm{u}_f$ is given as $\bm{u}_f = \langle \bm{u}_f\rangle + \bm{u}'_f$. It is notable that the Reynolds average of Eulerian quantities represents a volume average and a Reynolds average of a Lagrangian quantity represents an ensemble average, $\left\lbrack\sum_{t=t_0}^{t_f}\left(\sum_{i=1}^{N_p} \bm{u}^{(i,t)}_p dt\right)\right\rbrack/(t_f-t_0)$, where the superscripts $i,t$ denote the $i$-th particle at the $t$-th time step and the statistically stationary period is bounded by $t_0$ and $t_f$. 

Phase averaging (analogous to Favre averaging) is commonly used and results in phase-averaged equations, which enable multiphase turbulence modeling~\cite{beetham2021sparse}. These phase averages are denoted by $\langle \cdot \rangle_f$ and $\langle \cdot \rangle_p$ for fluid and particle phase averages, respectively. And are defined for some general quantity $a$ as $\langle a \rangle_p = \frac{\langle \varepsilon_p a \rangle}{\langle \varepsilon_p\rangle}$,
where $\langle \cdot \rangle$ indicates a Reynolds average. Because the particle-phase volume fraction, $\varepsilon_p$ is an Eulerian quantity, computed by projecting the Lagrangian particle volume to the Eulerian grid via trilinear interpolation and applying a Gaussian filter, when $a$ represents a Lagrangian quantity, it must also be recast as an Eulerian field. 

Thus, there also exist fluctuations about the particle phase average, defined as $\tilde{\bm{u}}_p^{''} = \tilde{\bm{u}}_p - \langle \tilde{\bm{u}}_p\rangle_p$. Here, the phase average of these fluctuations is null (i.e., $\langle \tilde{\bm{u}}_p^{''}\rangle_p \equiv 0$) but since $\tilde{\bm{u}}_p^{''} \neq \bm{u}_p'$, then $\langle \tilde{\bm{u}}_p^{''} \rangle \neq 0$. 

Owing to the two ways in which to view particle-phase information (either Lagrangian or Eulerian) and the two definitions of fluctuating particle phase velocity, the total fluctuating energy in the particle phase, e.g. the total granular energy, $\kappa_p$ (a Lagrangian quantity), can be decomposed as the sum of the particle phase turbulent kinetic energy (TKE), $k_p$ (an Eulerian quantity) and the phase-averaged granular temperature, $\Theta$, as 
\begin{equation}
\kappa_p = k_p  + \frac{3}{2}\langle \Theta\rangle_p.
\label{eq:Kappap_balance}
\end{equation}
Here, the total granular energy is a purely Lagrangian quantity defined as 
\begin{equation} 
\kappa_p = \frac{1}{2}\left\langle\bm{u}'_p\cdot \bm{u}'_p \right\rangle
\label{eq:kappaPdef}
\end{equation} 
and thus no information is lost via projection to the Eulerian grid. The particle phase TKE, $k_p$, is purely Eulerian and analogous to the Reynolds Stress tensor that $Ar$ises in single-phase turbulence. It is defined as 
\begin{align}
    k_p &= \frac{1}{2}\left\langle \tilde{\bm{u}}_p^{''} \cdot \tilde{\bm{u}}_p^{''}\right\rangle_{p} \label{eq:totflucPart}. 
\end{align}
Due to the projection and filtering required to compute $k_p$ and the notion that $\tilde{\bm{u}}_p^{''} \neq \bm{u}_p'$, this quantity is incapable of capturing the \emph{total granular energy}. Instead, $k_p$ represents the \emph{correlated} contribution to total granular energy. Given this, an \emph{uncorrelated} contribution to granular energy is required to resolve the proportion of granular energy that $k_p$ cannot resolve due to the filtering operation. This term is the particle phase averaged granular temperature, $\langle \Theta\rangle_p$~\cite{capecelatro2014investigating}.

This distinction is important in the context of coarse-grained methods, such as the EE or PIC formulations, where Lagrangian information is not available and must be reconstructed using the resolved portion of the granular energy ($k_p$) and the granular temperature, which requires a model. Further, both of these quantities are directly tied to the filter size used to cast Lagrangian information as an Eulerian field, and thus the relative contribution of each is also intimately connected to the resolution, or filter size, of the Eulerian-based approach chosen.

Given this, $\langle \Theta \rangle_p$ can be computed either from $\kappa_p$ and $k_p$ by way of Eq.~\ref{eq:Kappap_balance} or directly as 
\begin{equation}
\langle \Theta \rangle_p = \frac{1}{3} \widetilde{\delta\bm{u}_p \cdot \delta\bm{u}_p},
\end{equation} 
where $\widetilde{\cdot}$ represents the filtering operation and $\delta \bm{u}_p$ is the Lagrangian fluctuations relative to the Eulerian particle velocity, $\tilde{\bm{u}}_p$. 
These two approaches of computing $\langle \Theta \rangle_p$ are only equivalent when $\langle \bm{u}_p\rangle = \langle \tilde{\bm{u}}_p \rangle_p$ holds, however, the range of conditions studied in this work indicates that the validity of this argument depends upon filter size. Thus, we compute the granular temperature by way of Eq. \ref{eq:Kappap_balance}, using the known total granular energy and resolved contribution from TKE, $k_p$, which varies depending on filter size.

%It follows that the particle motion itself is resolved as systemic kinetic energy ($\kappa_{p}$) from the particles' direct interactions (pure Newtonian mechanics, as-seen in Eq. \ref{eq:KappaP}), however there is a contribution from both the diffused Lagrangian quantities to the Eulerian grid ($k_{p}$, defined in Eq. \ref{eq:KP}), and the granular temperature ($\langle \Theta_{p} \rangle$, the amount of unresolved kinetic energy that results from the inelastic collisions effectively acting as molecules colliding in a "traditional" gas)\cite{GOLDHIRSCH2008130}, giving the total granular energy, defined in Eq. \ref{eq:totgran}.\\

\section{Data and computational methodology}
\label{sec:method}
The data that underpins the analysis and modeling efforts presented in this work comes from a recently completed CFD-DEM simulation campaign of cluster-induced turbulence (CIT). This phenomenon has been previously described in the literature~\cite{capecelatro2014investigating} and represents the cascade of turbulent energy from small to large scales due to strong coupling between the phases. In particular, in gas-solid flows with sufficiently high mass loading ($\varphi = (\varepsilon_p\rho_p)/(\varepsilon_f\rho_f)\gg1$) particles spontaneously form coherent structures in the form of clusters, largely due to the mechanism of reduced drag. This heterogeneity has been shown to generate and sustain turbulence in the gas phase, as the name CIT suggests. 

To this end, the simulation campaign considered a triply periodic domain of prototypical gas and particles commonly found in industrial applications, such as circulating fluidized bed reactors (see Tab.~\ref{t.props}). As with previous similar studies, e.g., \citet{capecelatro2015on}, CIT is considered in a triply-periodic, rectangular domain domain with an aspect ratio of four. All simulation properties are shown in Tab.~\ref{t.props}. 

%In the CFD-DEM method, the particles are resolved individually, including all collisions with a linear-spring dashpot type contact model. NETL's open-source multiphase CFD code MFIX-Exa (\url{https://mfix.netl.doe.gov}) was used as the simulation code.
% In the gravity aligned direction ($y$), the length of the domain is $L_y = 2048d_p = 15.36$~cm. In the transverse directions, the domain width and depth are $L_x = L_z = 512d_p = 3.84$~cm. A uniform fluid mesh of size $dx = 2dp = 150$~micron is applied. 

% Lee/Sarah - feel free to delete below. Just left them here in case you would rather have them in table form rather than in text. 
%      domain width            & $L_x$       & $3.84$ cm \\
%      domain height           & $L_y$       & $15.36$ cm \\
%      domain depth            & $L_z$       & $3.84$ cm \\
%      fluid mesh size         & $dx$        & $150$ $\mu$m \\
%      
%      ``standard'' gravity    & $\left| \bm{g} \right|$ & $9.81$ m/s$^2$ \\
%      gravity coefficient     & $C_g$       & $1$ to $5$ \\ 
%      Archimedes number       & $Ar$        & $\approx 18$ to $\approx 90$ \\
%      particle count          & $N_p$       & $10270570$ to $410232797$ \\
%      mean concentration      & $\phi_0$    & $\approx 0.01$ to $\approx 0.4$ \\

%\wdf{I would just give one of $k_n$ or $dt_{coll}$ in the table below. I left them both b/c normally, as a reader possibly trying to replicate, I would prefer the exact value of $k_n$. But I realize this is pretty soft and could raise a flag for certain reviewers and it might be better to obfuscate that by just providing $dtcoll$.}

\begin{table}[!ht]
  \begin{center}
    \begin{tabular}{lcl l}
    \hline
    \hline
    \multicolumn{4}{c}{\textbf{Physical properties}}\\
    \hline 
      Gas density             & $\rho_g$    & $1.2$ & kg/m$^3$ \\
      Gas viscosity           & $\mu_g$     & $1.8 \times 10^{-5}$ & Pa$\cdot$s \\
      Particle diameter       & $d_p$       & $75$ & $\mu$m \\
      Particle density        & $\rho_p$    & $1200$ & kg/m$^3$ \\
      Gravity    & $\bm{g}$ & $(0,-9.81,0)$ & m/s$^2$ \\
      Gravity coefficient     & $C_g$       & $1$ to $5$ & [--]\\ 
      Collisional spring constant         & $k_n$       & $3.27$ & N/m \\
      Restitution coefficient & $e_{pp}$    & $0.90$ & [--] \\
      Interparticle friction coefficient    & $\mu_{pp}$  & $0.25$ & [--]\\
   %   collision duration      & $dt_{coll}$ & $\approx 2\times10^{-5}$ & s \\
   \hline
   \multicolumn{4}{c}{\textbf{Computational parameters}}\\
    \hline
      Domain size            & $(L_x \times L_y \times L_z)$       & $(3.84,15.36, 3.84)$ & cm \\
      Mesh size         & $\Delta x/d_p=\Delta y/d_p =\Delta z/d_p$        & $2.0$ & [--] \\
      Archimedes number       & $Ar$        & $18.38$ to $91.88$ & [--] \\
      Number of particles          & $N_p$       & $(10.2-410.2)\times10^6$ & [--]\\
      Mean concentration      & $\langle \varepsilon_p\rangle$    & $(0.01- 0.4)$ & [--]\\
      Mass loading & $\varphi$ & ( $10.10 - 666.67$) & [--]\\
      \hline 
      \hline
    \end{tabular}
  \end{center}
  \vspace{-1em}
  \caption{Physical properties of the CIT simulations and relevant simulation parameters.}
    \label{t.props}
\end{table}

The campaign consisted of 33 unique simulations sampled from a 2-D phase-space of mean solids concentration, ${\langle \varepsilon_p\rangle}$, and Archimedes number, $Ar = {\rho_g \Delta \rho \left|{\bf g}\right| d_p^3}/{\mu_g^2}$. Mean particle concentration ranged from 1\% to 40\% by changing the particle count from approximately 10 million to 400 million. The Archimedes number was adjusted by increasing the body force from one to five times standard gravity, ranging $Ar \approx 18$ to $Ar \approx 90$. The body force was balanced by a mean pressure gradient. In order to prevent the state from drifting due to numerical imprecision, an additional constraint on particle velocity was applied. To this end, the mean (global) particle velocity was normalized to zero to double precision at an interval of every fluid time step. For consistency, this was applied in all three directions giving $\langle \bm{u}_p\rangle = 0$. We note here that due to the transfer kernel and (if used) filtering, $\langle\bm{u}_p\rangle_{p}$ is negligibly small, though a few orders of magnitude larger than machine precision. 

Particles were initially seeded in a uniform hex-close packed lattice for all concentrations, rather than randomly distributed, for computational efficiency during initialization particularly at high volume fractions. Additionally, the particles were thermalized with an initial granular temperature of $\langle \Theta \rangle = 0.01$~m$^2$/s$^2$ for computational efficiency during startup. In other words, the particles were given a noisy initial velocity in an effort minimize the initial transient. 

Once a quasi-steady state was reached, which varies for each case and was determined ``by eye,'' AMReX-native plot files~\footnote{https://amrex-codes.github.io/amrex} were saved at regular intervals (every $\tau_p = {m_p}/{3\pi \mu_g d_p} = 0.02$~s), resulting in 100 snapshots of particle and fluid data during the statistically stationary period. All of the simulation snapshots, approximately 50~TB worth of binary plot file data, have been made openly available~\cite{fullmer2024machine}. 

The snapshots, i.e., AMReX plot files, were then post-processed with an AMReX-based application, \texttt{postmfix}, developed specifically for this purpose~\citep{fullmer2024machine}. In addition to the total particle fluctuating kinetic energy, $\kappa_p$ (Eq.~\ref{eq:Kappap_balance}), and all particle properties were computed by post-processing the simulation data with the same eight-cell, trilinear deposition scheme used in the simulation code to generate Eulerian fields from Lagrangian data. The continuous particle (i.e., solids) data can also be further smoothed by applying a Gaussian filter, which is a focus of this work. We note here that the original CFD-DEM simulations utilized an eight-cell deposition scheme in the numerical solution and the filtering used here is purely a post-processing step.

In this work, we employ a constant width filter, rather than a filter that varies based on the local volume fraction~\cite{esgandari2025grid}~\cite{capecelatro2014numerical}~\cite{capecelatro2015fluid} and ensures a fixed number of particles in the filtering region. The motivation for this choice is the fact that the filtered CFD-DEM data is analogous to the Eulerian mesh spacing for a coarse-grained CFD method, such as EE or PIC. In such approaches, the fluid mesh is typically uniform, or refined based on geometry not adaptively based on local volume fraction. Thus, filtering the highly resolved CFD-DEM data using a fixed filter width is most closely aligned with the fluid grid-based resolution chosen in the coarse grained approaches, approaches that require granular temperature models that should properly depend upon grid size.   

%\wdf{[Maybe here or on page 4 I think we need to defend using the constant filter width. We should cite Jesse's JFM 2014 and 2015 papers and the recent Schneiderbauer paper. I think the defense is that, to use this in something like PIC, a parcel has a given stat weight that is constant whether it is in a dense region or a dilute region. But I guess that means we should also introduce PIC in the Intro. I had that comment but there's not really a good palce for it. you wouldn't want to include it in the discussion of KT-TFM because it doesn't really have more closure laws like KT. so I guess it has to go into the filtered section below that. maybe need to introduce a filtered model too and say drag is the most important term and it's often applied to PIC as well which is usually treated like an ad-hoc filtered EL model.]}

\section{Clustering and granular temperature behavior}
\label{sec:results}
In this section, we examine both the clustering and granular temperature behavior across the parameter space studied and for a range of filter sizes. In particular, quantities of interest are computed for each of the 33 flow configurations (sampled from the $Ar$-$\langle \varepsilon_p\rangle$ parameter space discussed previously) and are subsequently filtered, using thirteen filter sizes that vary from 2 to 128 $d_p$ as $\delta_i = 2\cdot\delta_{i-1}$, with $\delta_0 = 2d_p$. This results in 429 data points.  

\begin{figure}[h!]
\centering
\subfigure[$\delta = 2 d_p$]{\label{fig:Dependence:a} \includegraphics[width=.3\textwidth]{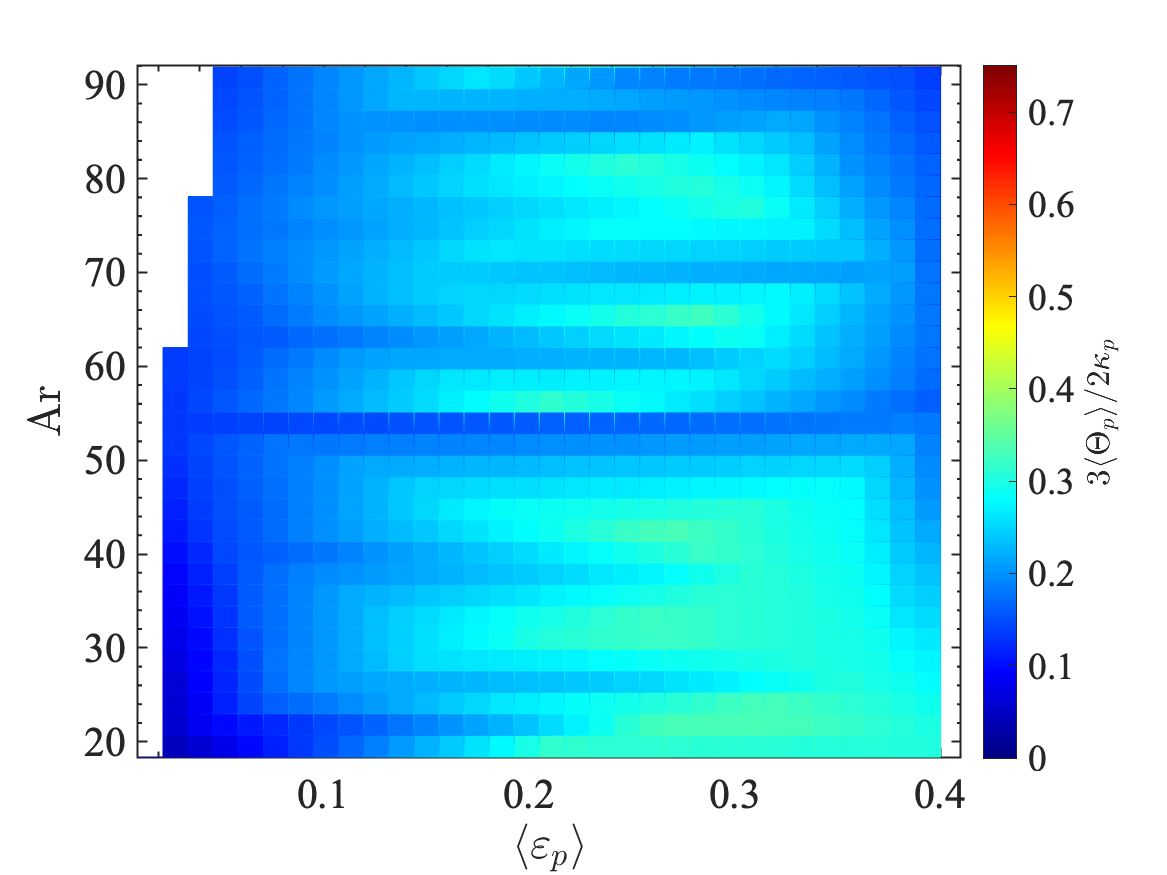}}
\subfigure[$\delta = 24 d_p$]{\label{fig:Dependence:b} \includegraphics[width=.3\textwidth]{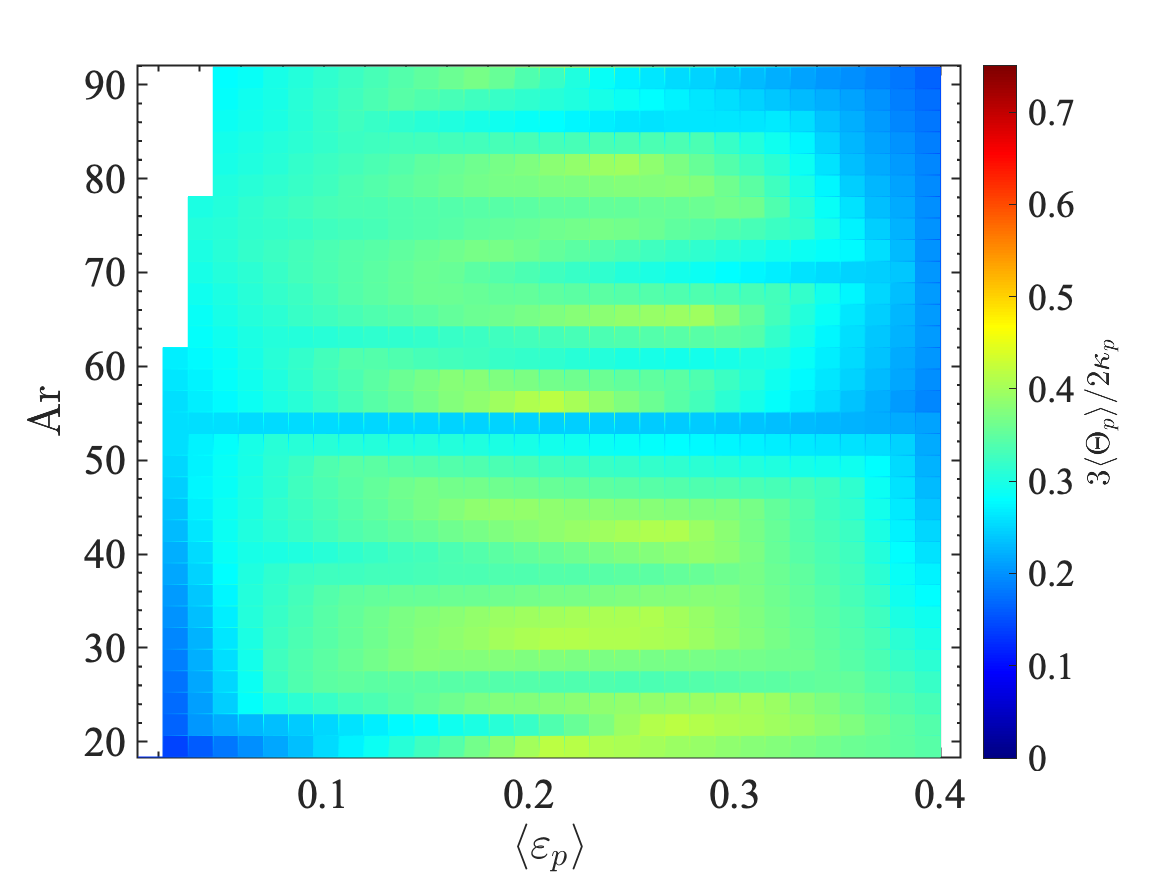}}
\subfigure[$\delta = 128 d_p$]{\label{fig:Dependence:c} \includegraphics[width=.3\textwidth]{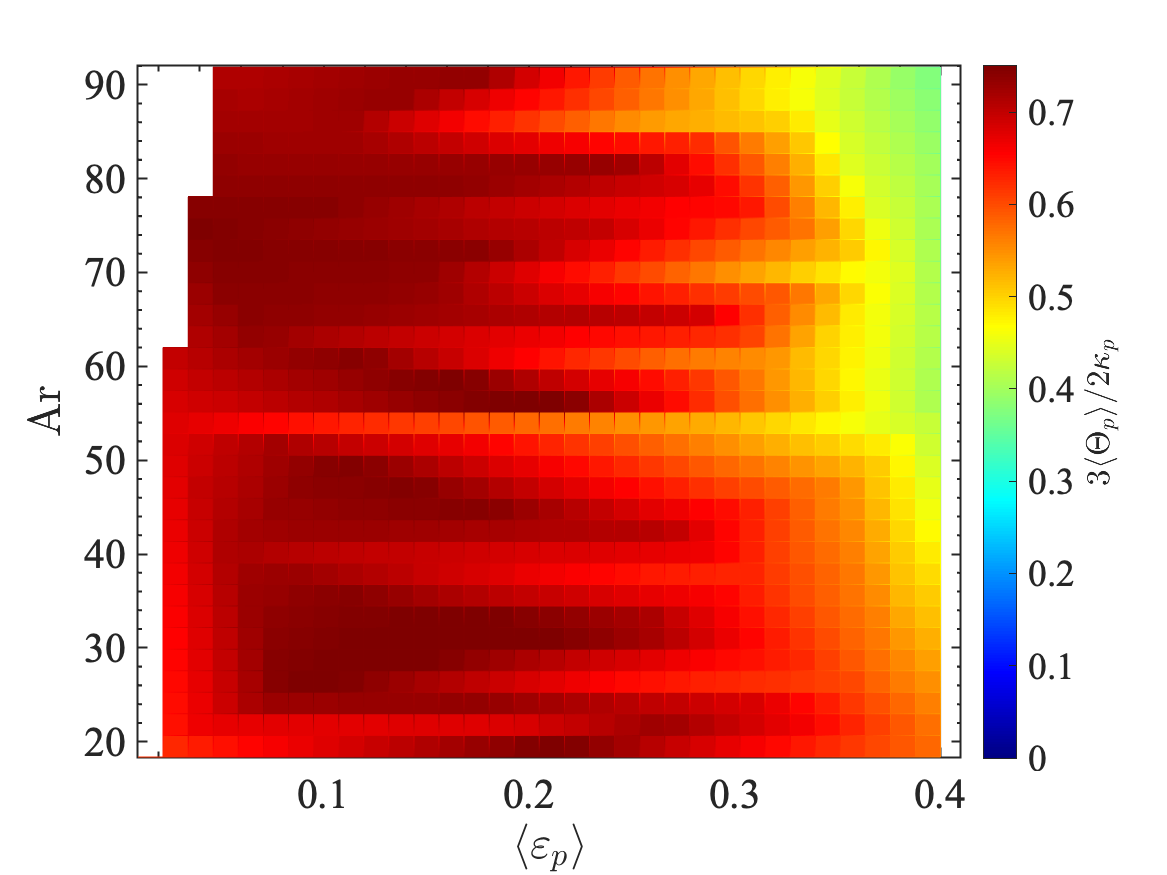}}
\caption{Normalized granular temperature dependence upon both $\langle \varepsilon_p\rangle$ and $Ar$ for three representative filter sizes. There is a clear dependence upon volume fraction, whereas the dependence on $Ar$ is weak and likely uncorrelated.}
\label{fig:Dependence}
\end{figure}

Since the simulation campaign perturbed two parameters (volume fraction and Archimedes number, $Ar$), it is useful to first consider how the balance of granular temperature changes with the normalized filter size, $\delta^{\star} = \delta/d_p$, as well as $\langle \varepsilon_p \rangle$ and $Ar$. As shown in Fig.~\ref{fig:Dependence} for three select filter sizes, there is a clear correlation between the normalized granular temperature ($3\langle \Theta_p\rangle/2\kappa_p$) and the mean volume fraction, $\langle \varepsilon_p \rangle$, however the dependence upon $Ar$ is weak.

\begin{figure}[h!]
\centering
  %  \subfigure[Full dataset]{\label{fig:a}  \includegraphics[width=10cm]{TKErunall.eps}} \\
\subfigure[Ar$=18.38, \; \langle \varepsilon_p\rangle = 0.010$]{\label{fig:ThetavsDeltaMulti:a} \includegraphics[width=.45\textwidth]{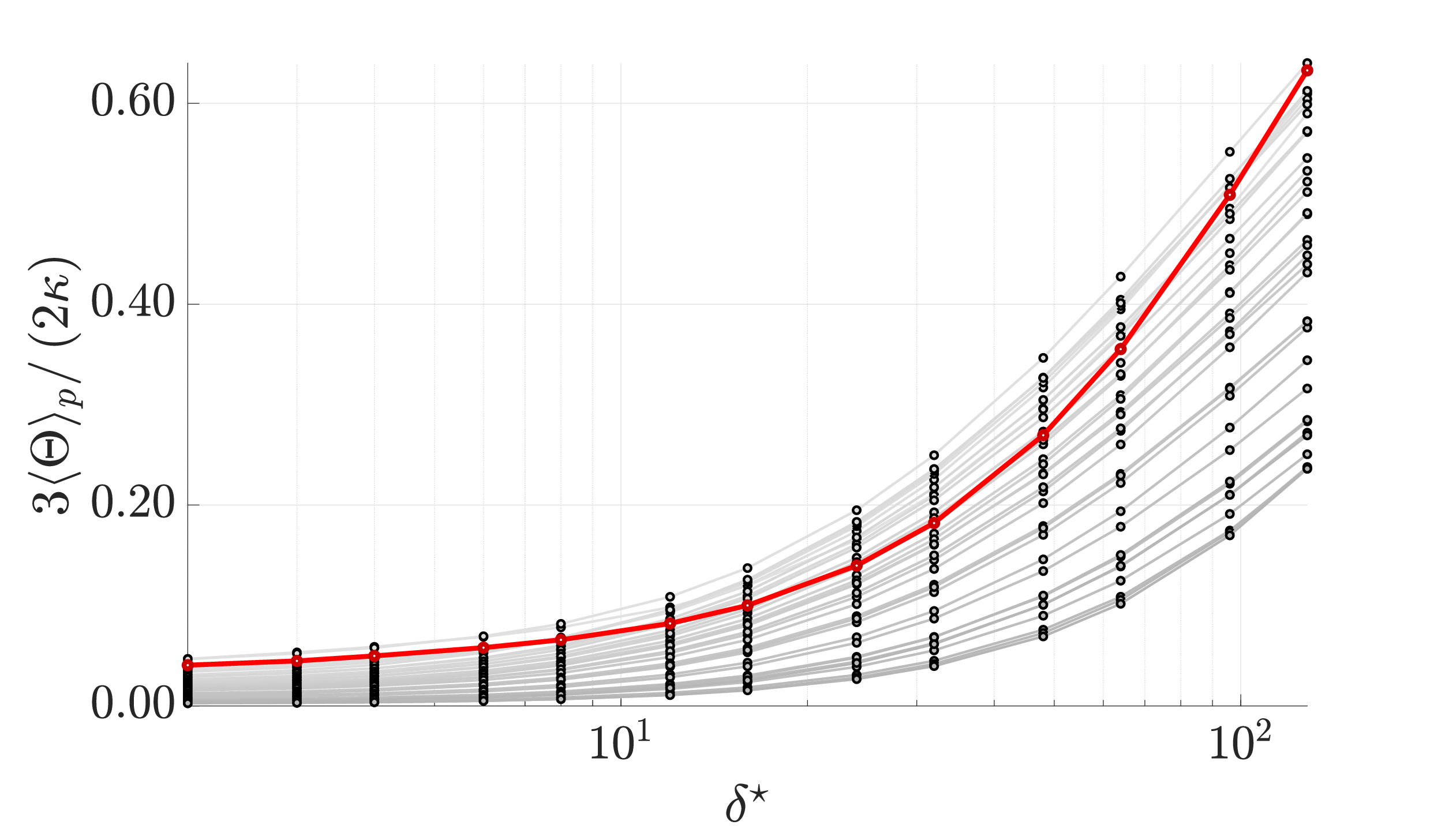}}
\subfigure[Ar$=91.88, \; \langle \varepsilon_p\rangle = 0.221$]{\label{fig:ThetavsDeltaMulti:b}\includegraphics[width=.45\textwidth]{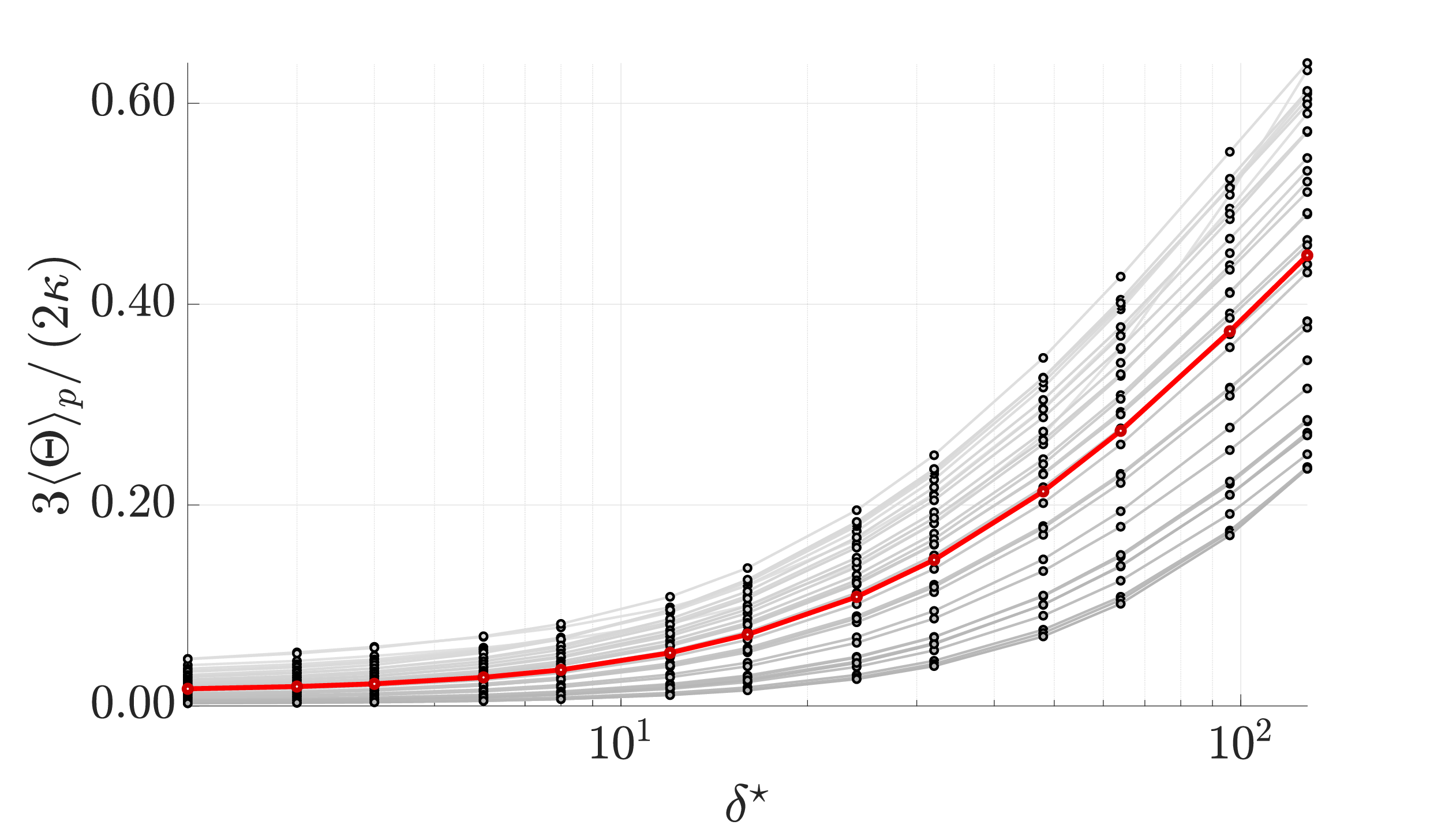}}\\
\subfigure[Ar$=59.10, \; \langle \varepsilon_p\rangle = 0.323$]{\label{fig:ThetavsDeltaMulti:c} \includegraphics[width=.45\textwidth]{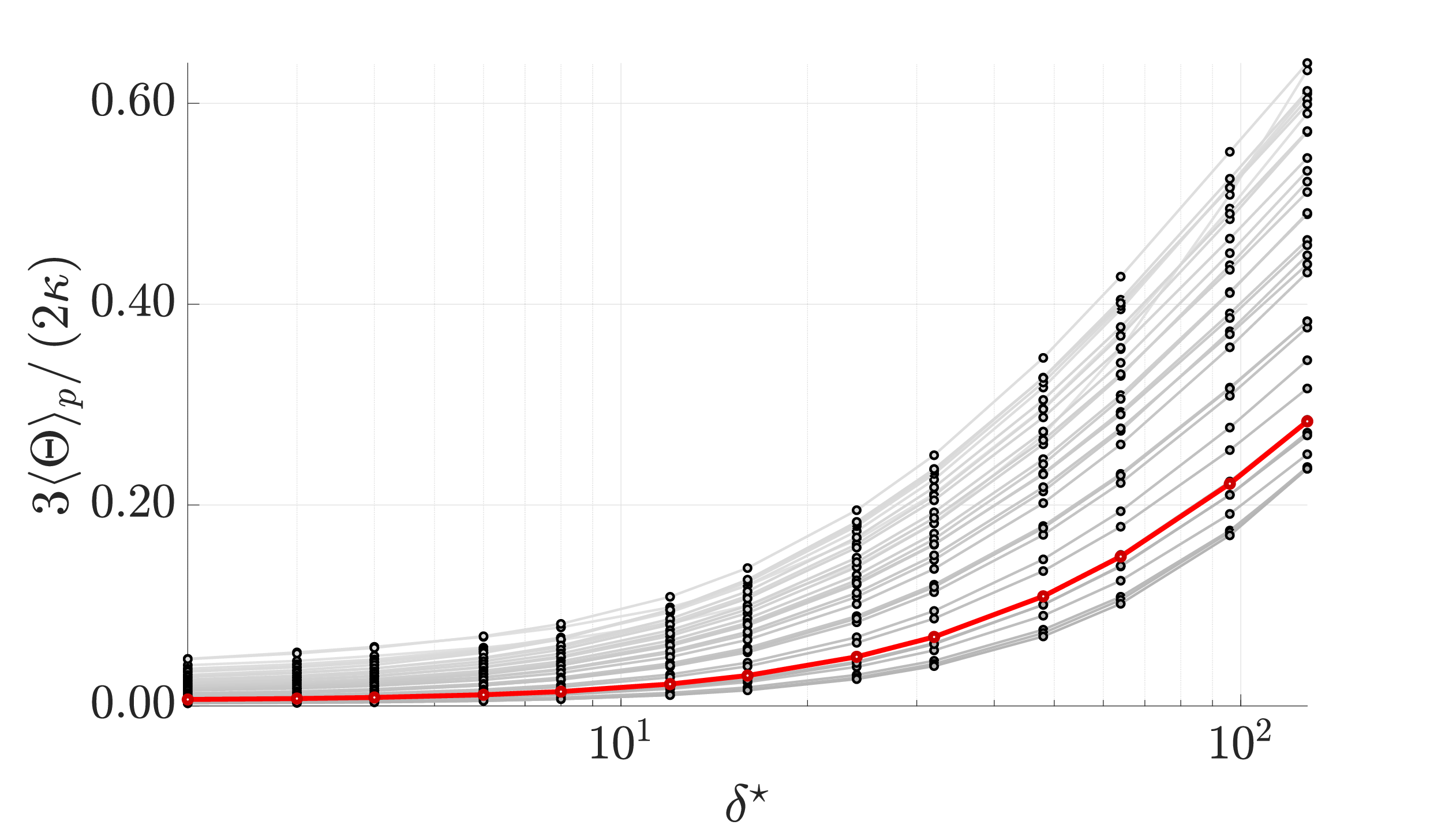}}
\subfigure[Ar$=55.13, \; \langle \varepsilon_p\rangle = 0.400$]{\label{fig:ThetavsDeltaMulti:d}\includegraphics[width=.45\textwidth]{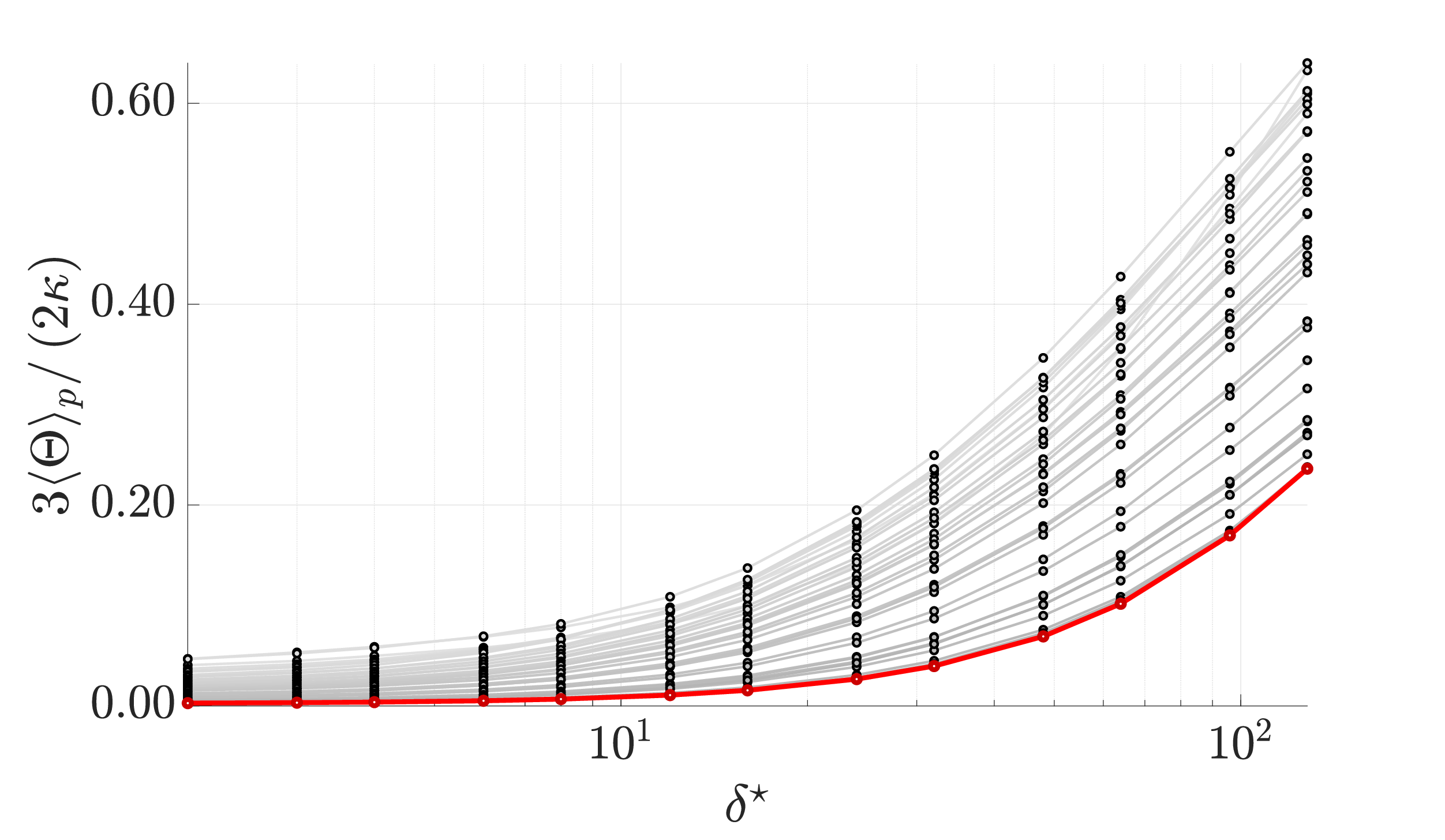}}
\vspace{-0.5em}
\caption{Granular temperature, normalized by the total granular energy increases with filter size. Here, the filter size, $\delta$ is normalized by $d_p$ ($\delta^{\star}=\delta/d_p$). The full dataset is displayed in each figure with gray circles and the dataset indicated in the caption is highlighted in red.}
\label{fig:ThetavsDeltaMulti}
\end{figure}

Next, we consider the dependence of the relative contribution of granular temperature to the total granular energy as a function of filter size. We observe that with increasing filter size the particle velocities become less correlated and, thus, the correlated contribution to $\kappa$, i.e., $k_p$, decreases and the uncorrelated contribution, $\Theta$, increases. This behavior is shown in Fig 2 of \citet{capecelatro2014investigating} and very similar behavior is observed in this data. We consider the ratio $3\langle \Theta_p\rangle/2\kappa_p$, which is shown in Fig.~\ref{fig:ThetavsDeltaMulti} as a function of filter size for all conditions with four cases highlighted in red. Here, we find that this dependence is monotonically but nonlinearly increasing. We also note that this trend that is consistent across the parameter space studied. In fact, it is more than consistent but nearly identical, i.e., sorting is independent of $\delta^\star$. In other words, most of the curves do not cross one another. However, this is not always the case, as highlighted in Fig.~\ref{fig:ThetavsDeltaMulti:a}. Fig.~\ref{fig:ThetavsDeltaMulti} also shows that, in general, the ratio decreases with increasing concentration, with the densest conditions lying at the bottom of the envelope of curves.

Because the degree of clustering plays an important role in granular temperature of an assembly of particles, it is also relevant to consider the degree of clustering present in the flow. Here, we make use of the parameter, $\mathcal{D}$, defined originally by \citet{Eaton1994} as 
\begin{equation}
\mathcal{D} = \frac{\left( \sqrt{\langle \varepsilon_p^{'2}\rangle-\langle \varepsilon_p^{'2}\rangle{\vert}_0}\right)}{\langle \varepsilon_p \rangle}\approx \frac{ \sqrt{\langle \varepsilon_p^{'2}\rangle}}{\langle \varepsilon_p \rangle}
\label{eq:clustDeg}
\end{equation}
where $\sqrt{\langle \varepsilon_p^{'2}\rangle}$ is the variance of the particle volume fraction of the fully-developed configurations and $\sqrt{\langle \varepsilon_p^{'2}\rangle}\vert_0$ is the variance of the volume fraction of an \emph{uncorrelated} assembly of particles (i.e., the initial configuration in this simulation campaign), which is assumed to be null. Given this definition, $\mathcal{D}$ will become increasingly large with higher deviations from an uncorrelated state, thus indicating the degree of clustering. 

\begin{figure}[h!]
\centering
    \subfigure[The relationship between $\mathcal{D}$ and $\langle \varepsilon_p \rangle$ in relation to filter size.]{\label{fig:var_vfmodel:a} \includegraphics[width=.32\textwidth]{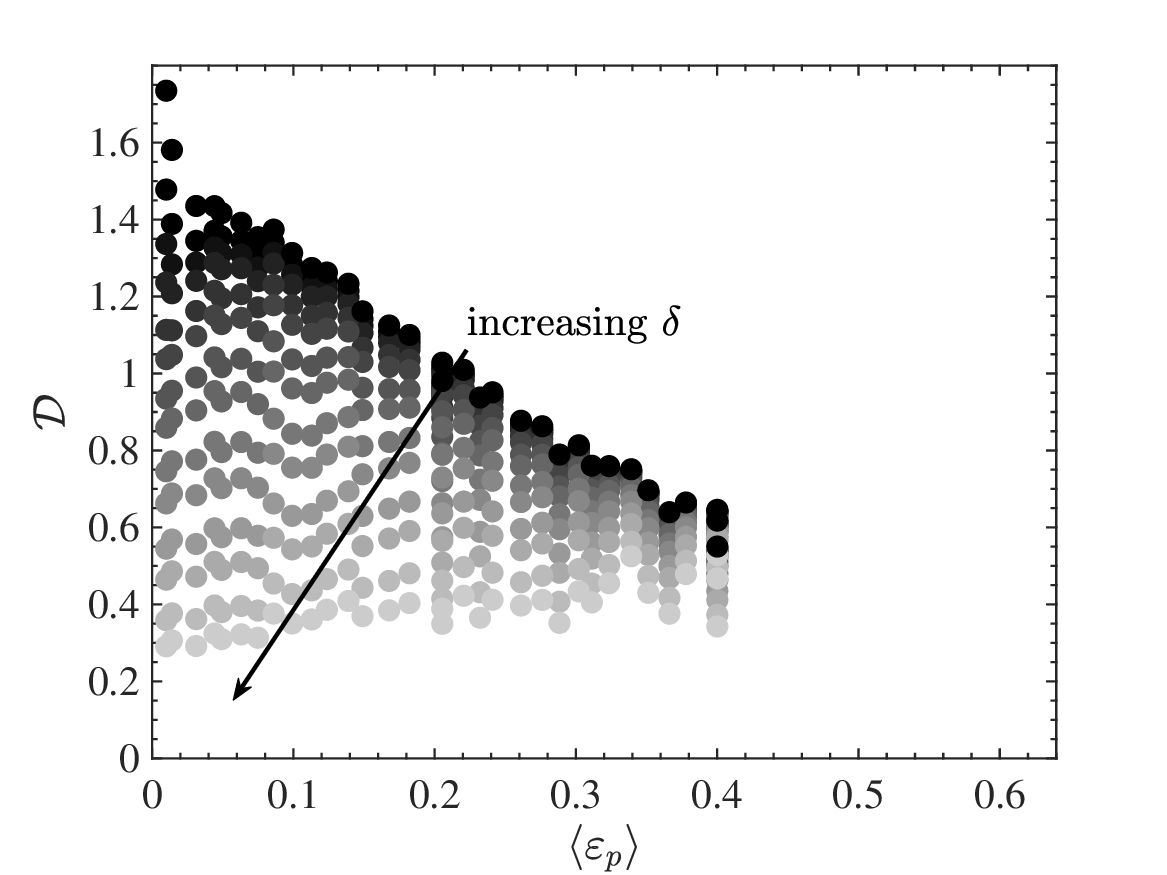}}
    \subfigure[$\sqrt{\langle \varepsilon^{\prime 2}_p\rangle}$ vs $\langle \varepsilon_p \rangle$]{\label{fig:var_vfmodel:b} \includegraphics[width=.32\textwidth]{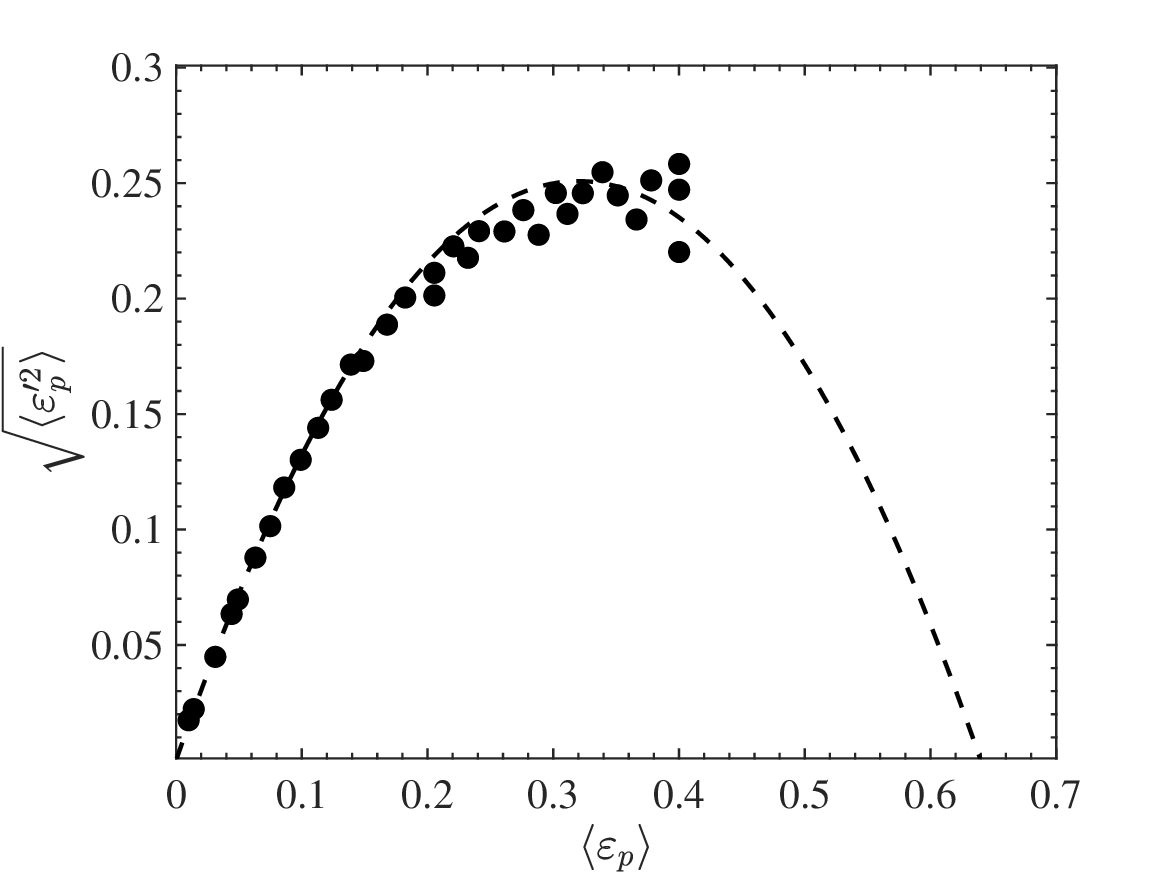}}
    \subfigure[$\mathcal{D}$ vs $\langle \varepsilon_p\rangle$]{\label{fig:var_vfmodel:c}\includegraphics[width=.32\textwidth]{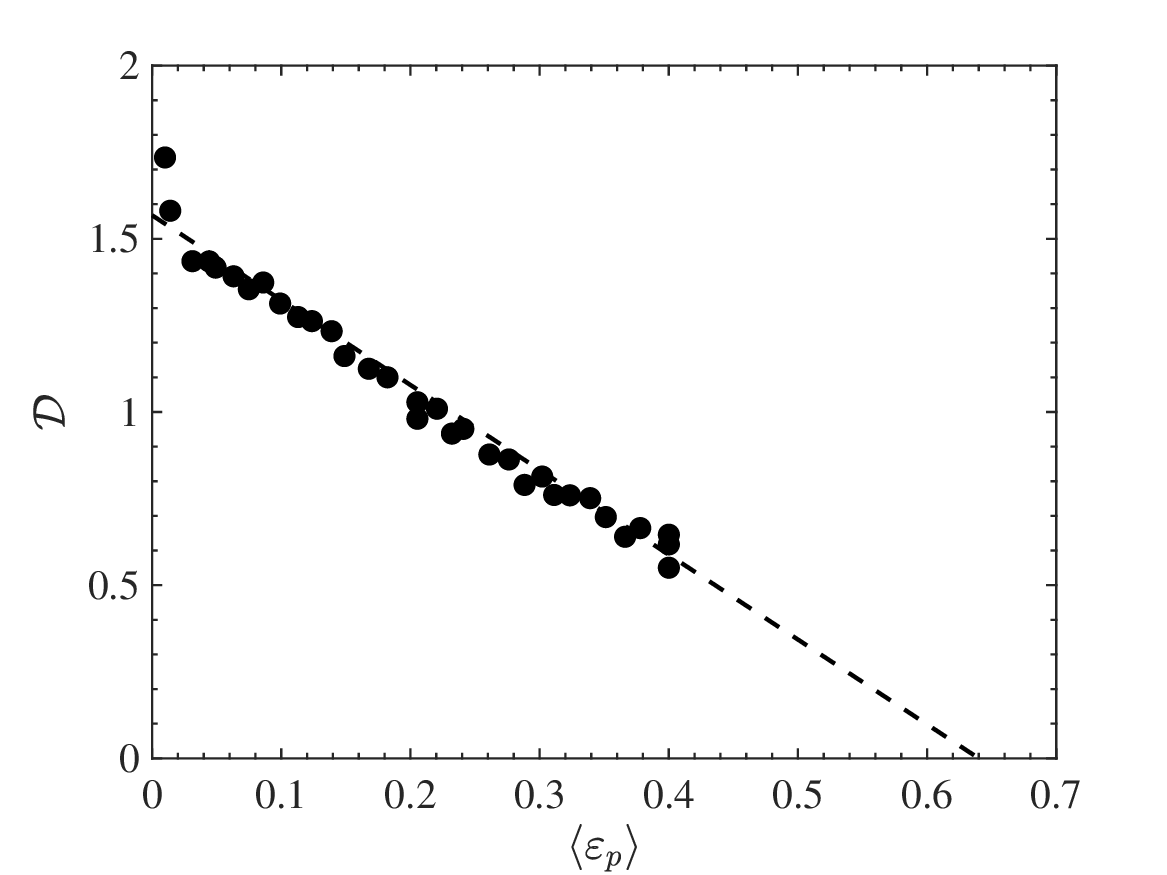}}
\caption{Comparison of highly resolved data (symbols) and the model proposed in Eq.~\ref{eq:ModelVFvar} (dashed line).}
\label{fig:var_vfmodel}
\end{figure}

In the work of \citet{issangya2000further}, a model for the variance of particle volume fraction was formulated as 
\begin{align}
 %  \sqrt{\langle \varepsilon_p^{\prime 2}\rangle}  &=  1.584(\langle\varepsilon_p\rangle((1-\varepsilon_{mf})-\langle\varepsilon_p\rangle) \\ \nonumber
  \mathcal{D}&= 1.584\left\lbrack(1-\varepsilon_{mf})-\langle\varepsilon_p\rangle\right\rbrack \\
  &= 1.584(\varepsilon_{mp}-\langle\varepsilon_p\rangle)
\label{eq:issangyaVF}
\end{align}
where $\varepsilon_{mf}$ and $\varepsilon_{mp}$ represent the theoretical maximum volume fractions in the fluid and particle phases, respectively. In \citet{issangya2000further}, $\varepsilon_{mf}$ was measured to be 0.55 which is a typical fluffed packing in practice, i.e., in a real system. However, the underlying data in this case comes from CFD-DEM simulations in which the particles are perfectly spherical, monodisperse and smooth. Therefore, we have changed $\varepsilon_{mf}$ to the spherical random close-packed limit of 0.64 in our proposed improved model in addition to fitting the leading coefficient to the most highly resolved data in this study (corresponding to a filter size of $\delta = 2 d_p$). Thus, we propose an adjusted model, given by
%however, it is possible for a granular assembly to attain a mean volume fraction of 0.64 in the case of random close packed particles. Thus, using this value would predict \emph{negative} values for the variance of volume fraction for physically realizable systems. Instead, when the volume fraction approaches the random close packed limit, assumed in this work to be 0.64, the degree of clustering should be null. 
\begin{align}
%\sqrt{\langle \varepsilon_p^{\prime 2}\rangle} &= 2.4\langle\varepsilon_p\rangle(\varepsilon_{mp}-\langle\varepsilon_p\rangle) \nonumber \\
\mathcal{D} &= 2.4(\varepsilon_{mp}-\langle\varepsilon_p\rangle) \\
&= 2.4(0.64-\langle\varepsilon_p\rangle)\label{eq:ModelVFvar} 
\end{align}
This model and the data that it was based upon is shown in Fig.~\ref{fig:var_vfmodel:b} and Fig.~\ref{fig:var_vfmodel:c}. 

After applying increasingly large filters, we observe a flattening of this curve, indicative of the loss of information at the sub-filter scale, as expected (see Fig.~\ref{fig:var_vfmodel:a}). Since this quantity is unclosed no matter the filter size, only a single model is needed. Thus, we use the most highly-resolved data, assumed to be the `ground truth', to inform the adjusted leading coefficient in the model described by Eq.~\ref{eq:ModelVFvar} (shown in Figs.~\ref{fig:var_vfmodel:b} and \ref{fig:var_vfmodel:c}).  

\begin{figure}[h!]
    \centering
    \subfigure[$\delta^{\star}$ for $(\langle \varepsilon_{p}\rangle, {Ar})=(0.04, 91.8$)]{\label{fig:normfluc:a} \includegraphics[height=.3\textwidth]{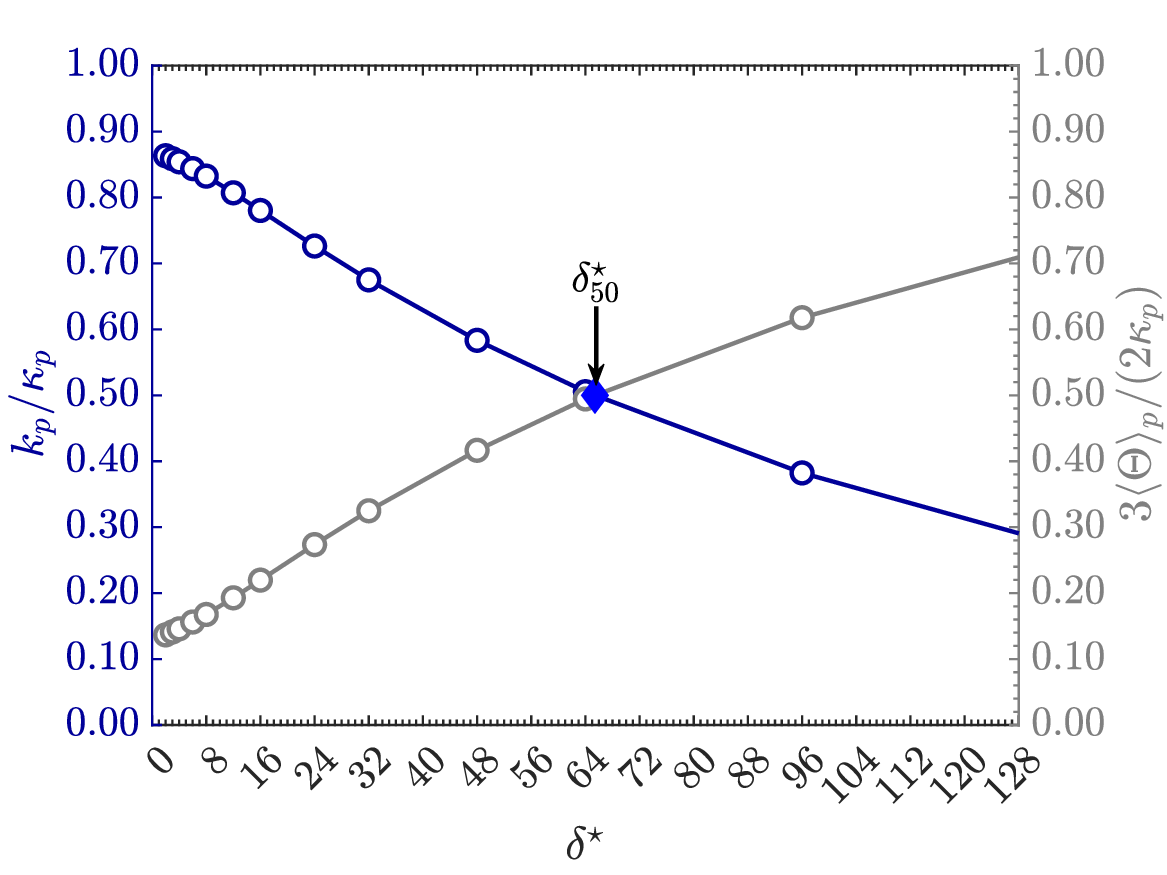}} 
    \subfigure[The relationship between $\delta^{\star}_{50}$ and $\langle\varepsilon_p\rangle$]{\label{fig:normfluc:b} \includegraphics[height = 0.3\textwidth]{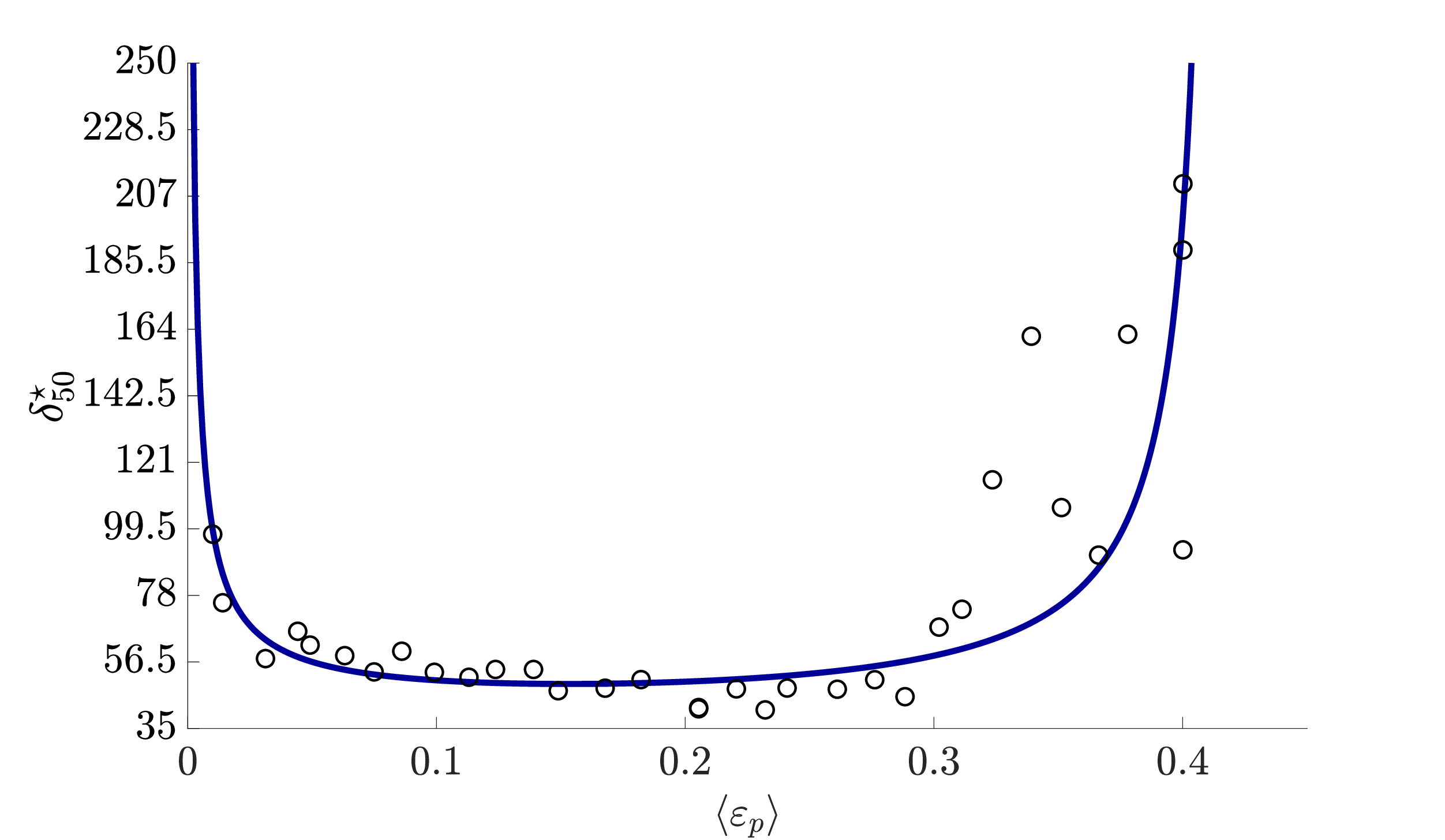}}
    \caption{The filter size at which granular temperature dominates the total granular energy is consistent for mid-range volume fractions, but increases asymptotically for dilute and dense assemblies.}
    \label{fig:normfluc}
\end{figure}

Finally, since granular temperature represents the \emph{unresolved} contribution to the total granular energy, it is useful to understand at what filter size this term dominates the balance. In other words, at what filter size does $3\langle \Theta_p \rangle/2 \ge k_p$? This can be determined by identifying the intersection of these two curves as a function of the normalized filter size, $\delta^\star$, for each configuration. An exemplary case is shown in Fig.~\ref{fig:normfluc}. Here, we denote this critical filter size as $\delta^{\star}_{50}$. 

Given this, it is interesting to note that the value of $\delta^{\star}_{50}$ is relatively consistent in the `mid range' of volume fractions considered (i.e., $0.05\leq \langle\varepsilon_p\rangle\leq0.25$). In this regime, the granular temperature becomes dominant at filter sizes larger than $\approx 60 \;d_p$. Interestingly, the granular temperature does not become dominant until the filter size is very large in the dilute and dense limits and for differing reasons. For a constant filter size in dilute suspensions, the filter size must get increasingly large in order to average a sufficient number of particles to attain a granular temperature that is half of the total fluctuating kinetic energy. On the other hand, for dense suspension, the heterogeneous state is one of dense regions near maximum packing and dilute regions (bubbles). Hence, while a filter located in a dense region may have an inordinate number of particles in it, it must grow large enough to the encompass macroscopic heterogeneity. % I removed the WDF directive
%We postulate that this is likely due to the fact that in these limits, \textcolor{red}{I forgot why we observe this but recall having a good reasoning for it... }

\subsection{A filter-dependent model for granular temperature}
\label{sec:TKE}

In a prior study by \citet{tang2016direct}, a model for granular temperature was postulated through the definition of a granular temperature-based Reynolds number, defined as 
\begin{equation}
\text{Re}_{\Theta}=\frac{\rho_{g}\text{d}_{p}\sqrt{\langle\Theta\rangle_{p}}}{\mu_{g}}.
    \label{eq:retheta}
\end{equation}
Based on a PR-DNS study of gas-solid flow with elastic collisions, a model for Re$_{\Theta}$ was formulated as 
\begin{equation}
    Re_{\Theta}\left(\text{Re},\frac{\rho_p}{\rho_g}\right) = 2.108 Re_p^{0.85}\sqrt{\frac{\rho_p}{\rho_g}}
    \label{eq:tangmodel}
\end{equation}
with the particle Reynolds number defined as 
\begin{equation}
    \text{Re}_{p}=\frac{\rho_{g}\text{d}_{p}\left(1-\varepsilon_{p}\right)\left\vert {\langle \bm{u}_{p} \rangle} - {\langle \bm{u}_{g}\rangle}\right\vert}{\mu_{g}}.
    \label{eq:rep}
\end{equation}

%An initial investigation by \citet{raval2024granular} showed that this model, while inaccurate, correctly postulated the functional form of the the dependence of Re$_{\Theta}$ on Re$_p$ for the data collected in this simulation campaign. This is particularly interesting considering the known limitations of PR-DNS studies--they are computationally constrained to small domain sizes and relatively few numbers of particles and are thus unable to capture mesoscale behavior such as CIT. 

%Thus, the remainder of this work focuses on adapting a model of the same functional form as proposed by \citet{tang2016direct} for the filter-dependent data collected. In other words, we begin with the generalized expression for Re$_{\Theta}$, 

In a preliminary study \citet{raval2024granular}, it was found that the functional form proposed by \citet{tang2016direct} held even for this substantially larger CIT data. However, the leading coefficient and exponent needed to be adjusted. It was also found that the adjustment was different depending on whether or not a filter was applied. Here, we consider a wide range of filter sizes, mentioned previously, and attempt to fit these coefficients as a continuous function. Specifically, we seek to fit

\begin{equation}
    Re_{\Theta}(\delta^{\star}) = A_{\Theta}(\delta^{\star}) Re_p^{B_{\Theta}(\delta^{\star})}\sqrt{\frac{\rho_p}{\rho_g}},
    \label{eq:powReg}
\end{equation}
where the coefficients $A_{\Theta}$ and $B_{\Theta}$ both depend upon the filter size, $\delta^{\star}$. 

\begin{figure}[h!]
\centering
\subfigure[Data with filter size $\delta^{\star}=8$]{\label{fig:ReTa} \includegraphics[width=0.48\columnwidth]{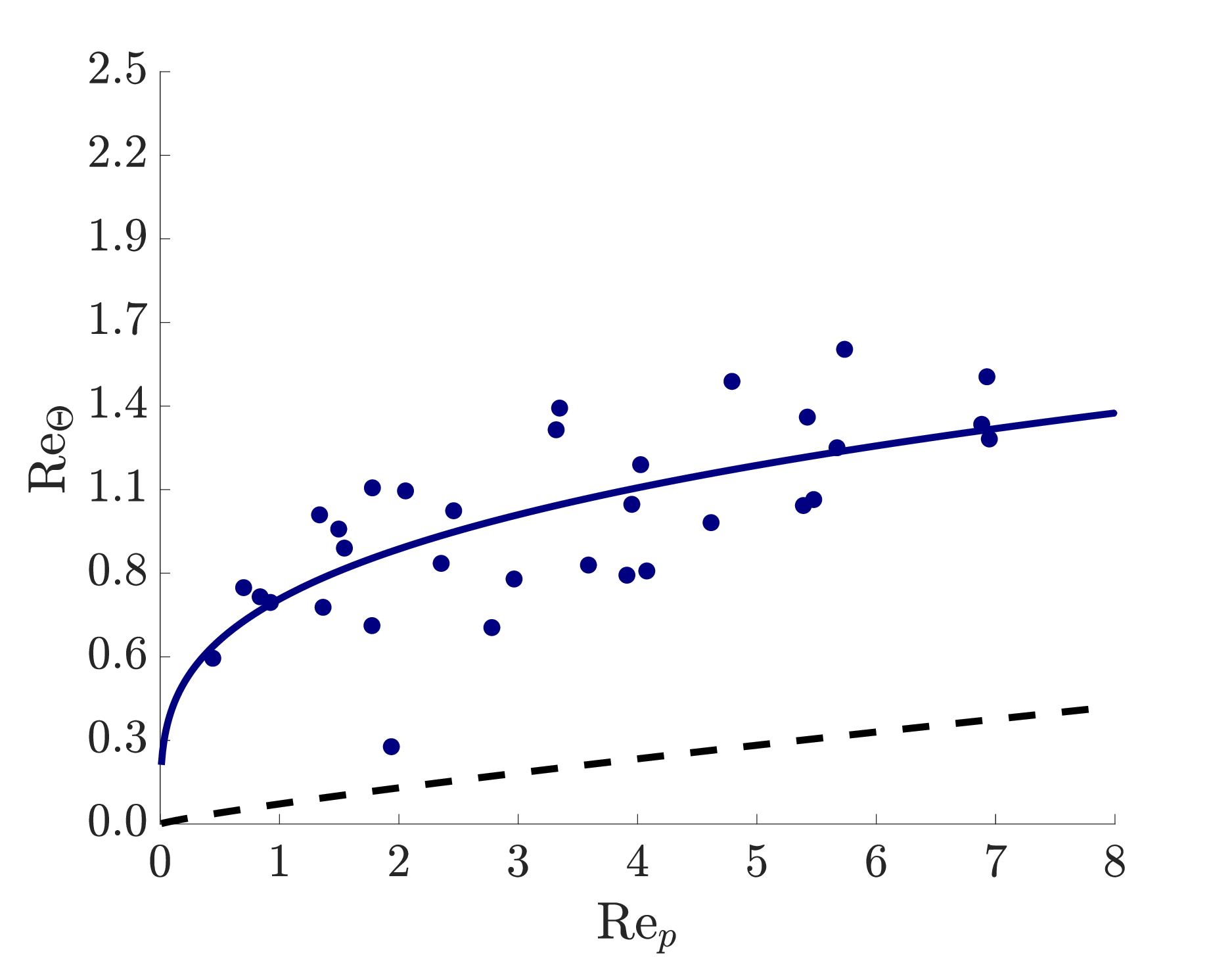}}
\subfigure[Data with filter size $\delta^{\star}=96$]{\label{fig:ReTb} \includegraphics[width=0.48\columnwidth]{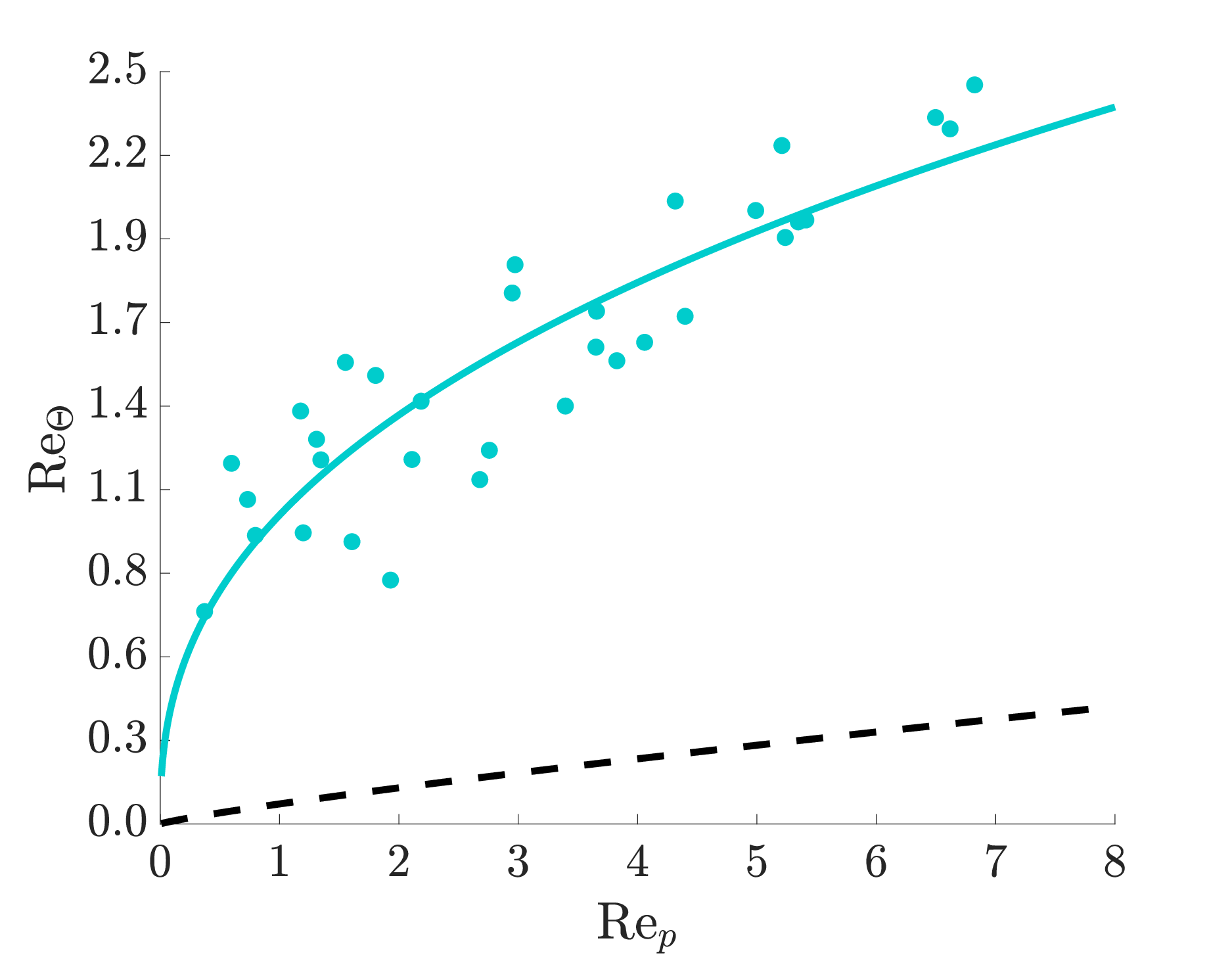}}
\subfigure[Data with filter size $\delta^{\star}=96$]{\label{fig:ReTSurface} \includegraphics[height=0.3\textwidth]{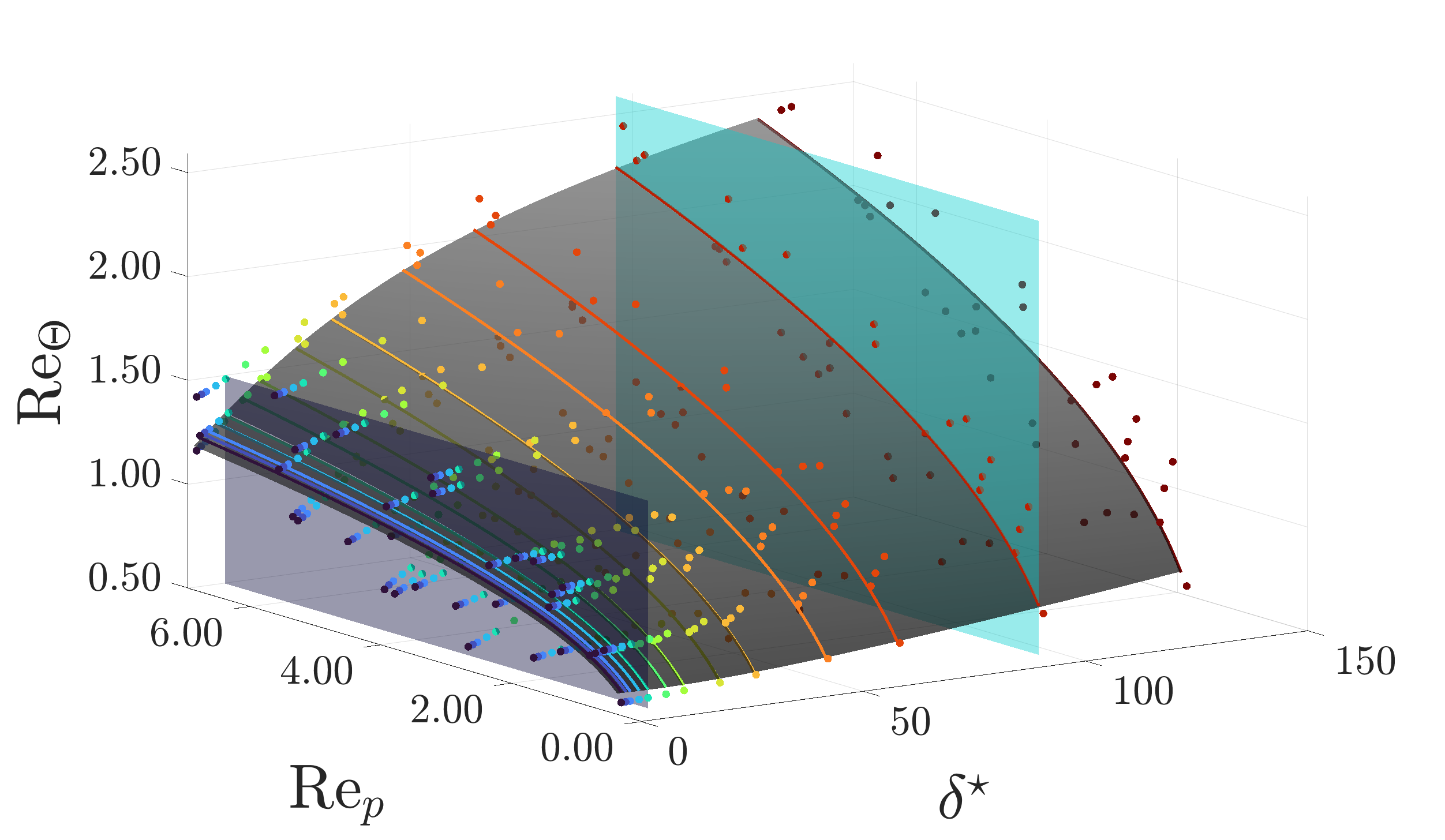} \hspace{-3.5em}\includegraphics[height=0.28\textwidth]{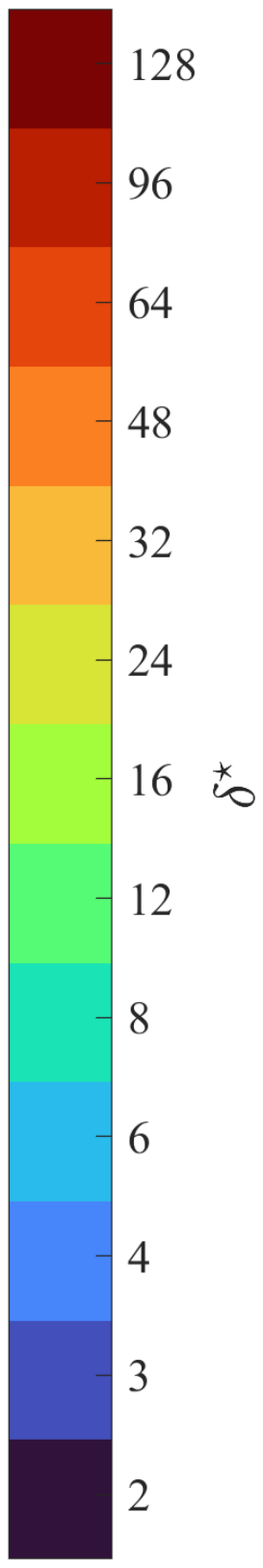}}
\caption{CFD-DEM data (symbols) are compared against the original model of \citet{tang2016direct}(dashed lines) and the new model with coefficients $A$ and $B$ fitted to the data (solid lines) at two exemplary filter sizes (a) and (b). The resulting model of Eq.~\ref{eq:powReg} is shown as a surface in (c). Note, the representative planes shown in (a) and (b) are called out in the three dimensional plot of (c).}
\label{fig:OldandNewModels}
\end{figure}

To carry this out, Re$_p$ and Re$_{\Theta}$ were computed based on the resulting data from the configurations studied. For each configuration, these quantities were computed over the range of filter widths studied (see two exemplary cases in Fig.~\ref{fig:ReTa} and \ref{fig:ReTb}). Then, for each filter size, constant values for $A_{\Theta}$ and $B_{\Theta}$ were determined by regressing the CFD-DEM data against the functional form of the model in Eq.~\ref{eq:powReg}. 

\begin{figure}[h!]
    \centering
      {\includegraphics[width=0.65\linewidth]{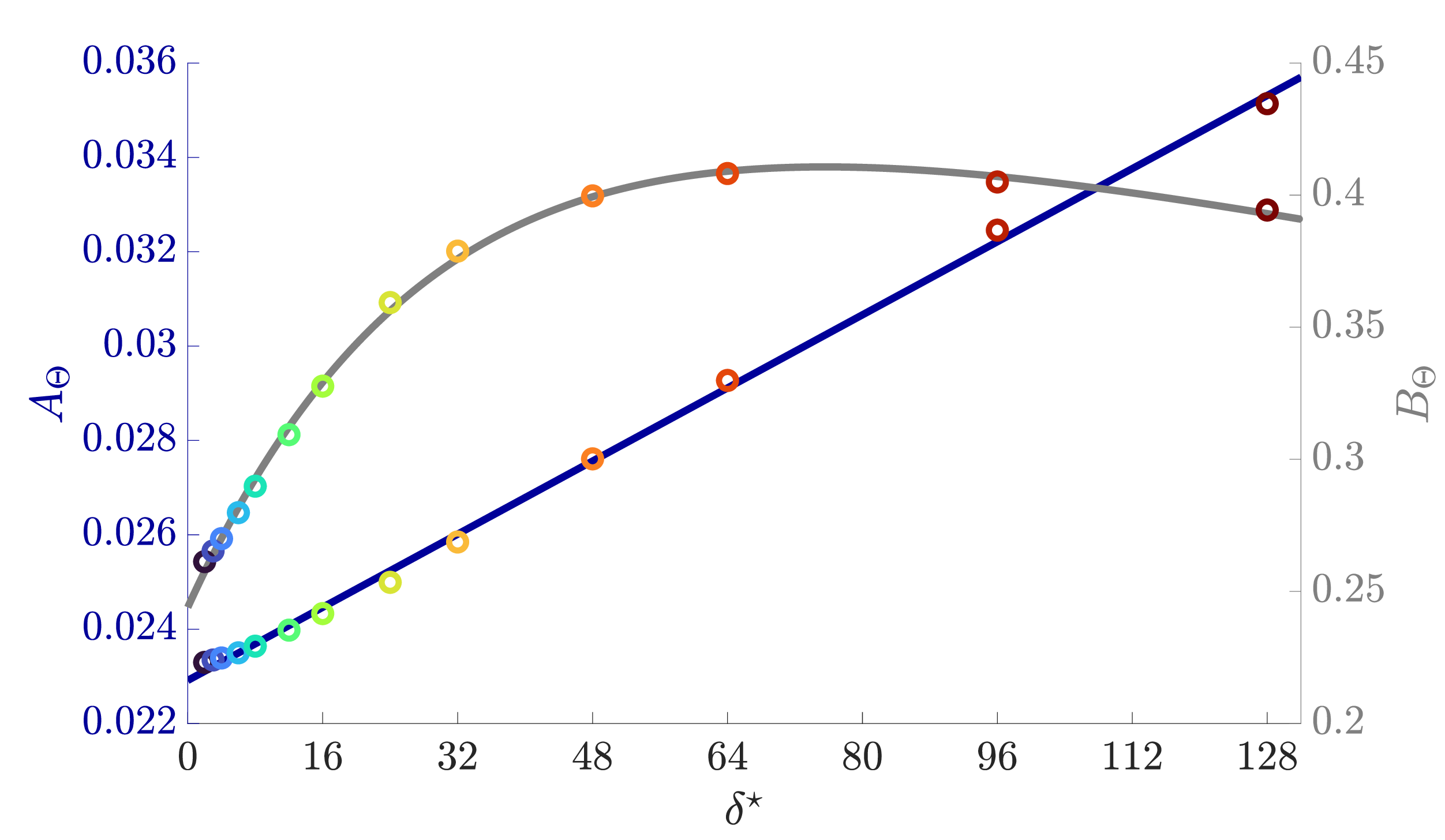}} 
    \caption{The fitted coefficients $A_{\Theta}$ and $B_{\Theta}$ in Eq.~\ref{eq:retheta} are dependent upon filter size. Note, each marker represents the fitted coefficient based on the 33 data points at each filter size.}
    \label{fig:AandB}
\end{figure}

This process resulted in unique values for $A_{\Theta}$ and $B_{\Theta}$ for each filter size (e.g., unique values for $A_{\Theta}$ and $B_{\Theta}$ describe the curves in Figs.~\ref{fig:ReTa} and \ref{fig:ReTb}). Thus, in order to collapse both into coefficients into functions dependent upon $\delta^{\star}$ rather than discrete values for each filter size, we then considered the dependence of $A_{\Theta}$ and $B_{\Theta}$ on filter size, $\delta^{\star}$ (see Fig.~\ref{fig:AandB}). 

Here, we note that $A_{\Theta}$ has a linear relationship on $\delta^{\star}$, given as 
\begin{equation} 
A_{\Theta} = 0.03(0.003\;\delta^{\star}+0.72).
\end{equation}

The dependence of $B_{\Theta}$ is slightly more complex, and is given as 
\begin{equation}
    B_{\Theta} = 0.49\exp\left(-0.0016\delta^{\star}\right) - 0.24\exp\left(-0.0316\delta^{\star}\right).
\end{equation}

%\wdf{[we need to note that A and B do not approach Tang's values as $\delta^\star \to 0$ because a zero filter size in this large problem size is not the same as the small problem size of DNS, ie we still have heterogeneous CIT state.]}

%\wdf{[I'm not sure putting A and B into the eq. and providing it as (4.10) really adds anything. if we want to re-emphasize it, maybe just give it all again in the conclusions:}

%We have proposed a large-scale granular temperature model for CIT of the form 
%\begin{equation}
%    Re_{\Theta}(\delta^{\star}) = A_{\Theta}(\delta^{\star}) Re_p^{B_{\Theta}(\delta^{\star})}\sqrt{\frac{\rho_p}{\rho_g}},
%\end{equation}
%with 
%\begin{equation} 
%A_{\Theta} = 0.03(0.004\;\delta^{\star}+1).
%\end{equation}
%and 
%\begin{equation}
%    B_{\Theta} = 0.49\exp^{-0.0016\delta^{\star}} - %0.24\exp^{-0.0316\delta^{\star}}.
%\end{equation}
%functions of the filter size $\delta^\star = \delta/d_p$.

\begin{figure}[h!]
\centering
 \subfigure[CFD-DEM results]{\label{fig:2Da} \includegraphics[width=.32\textwidth]{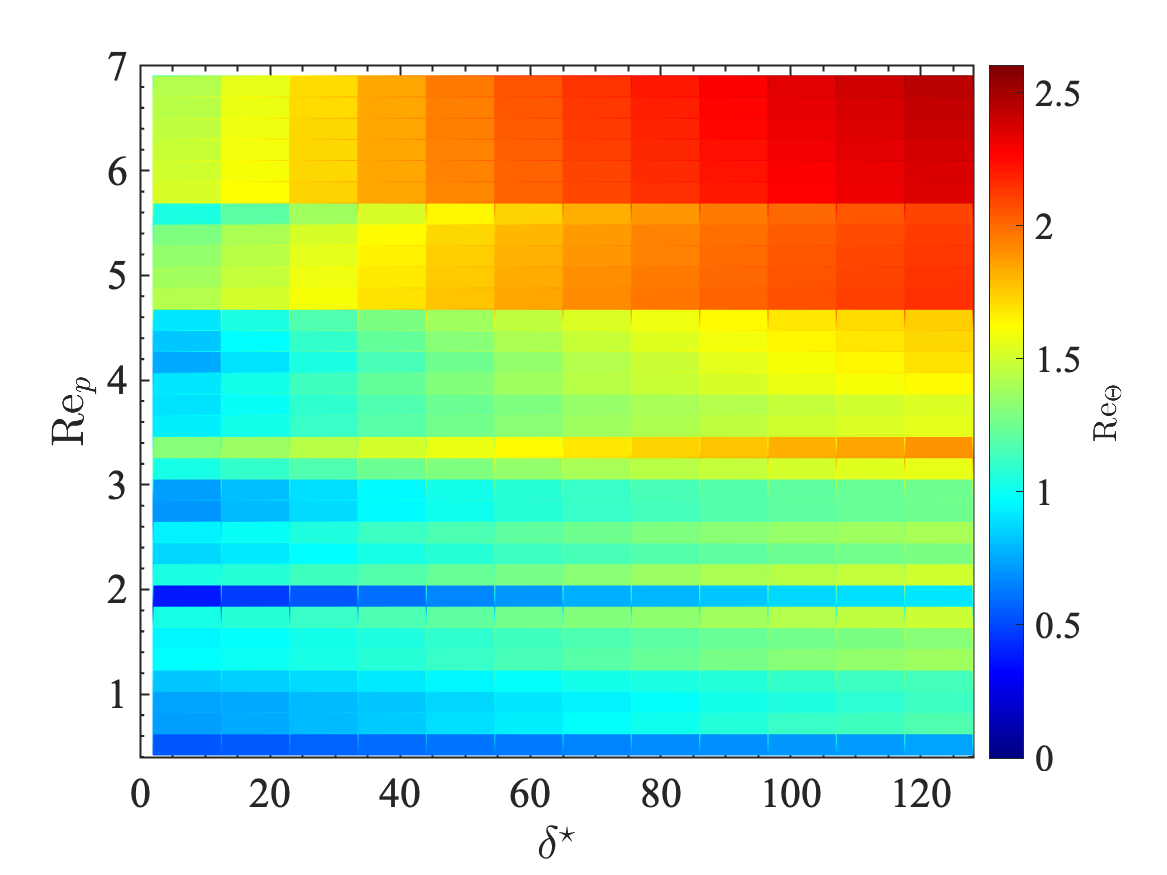}}
    \subfigure[Model prediction ( Eq.~\ref{eq:powReg})]{\label{fig:2Db} \includegraphics[width=.32\textwidth]{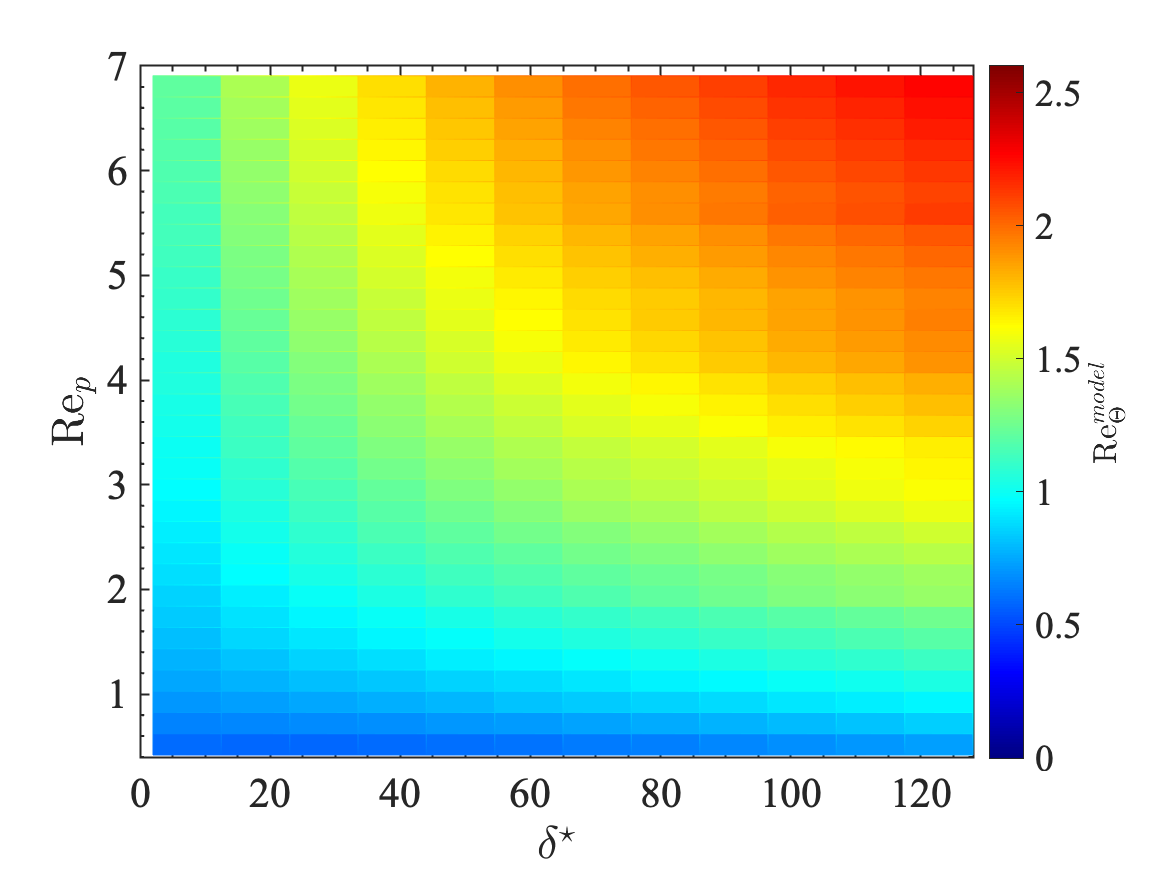}}
\subfigure[Model error, (mean: 20\%)]{\label{fig:2Dc}\includegraphics[width=.32\textwidth]{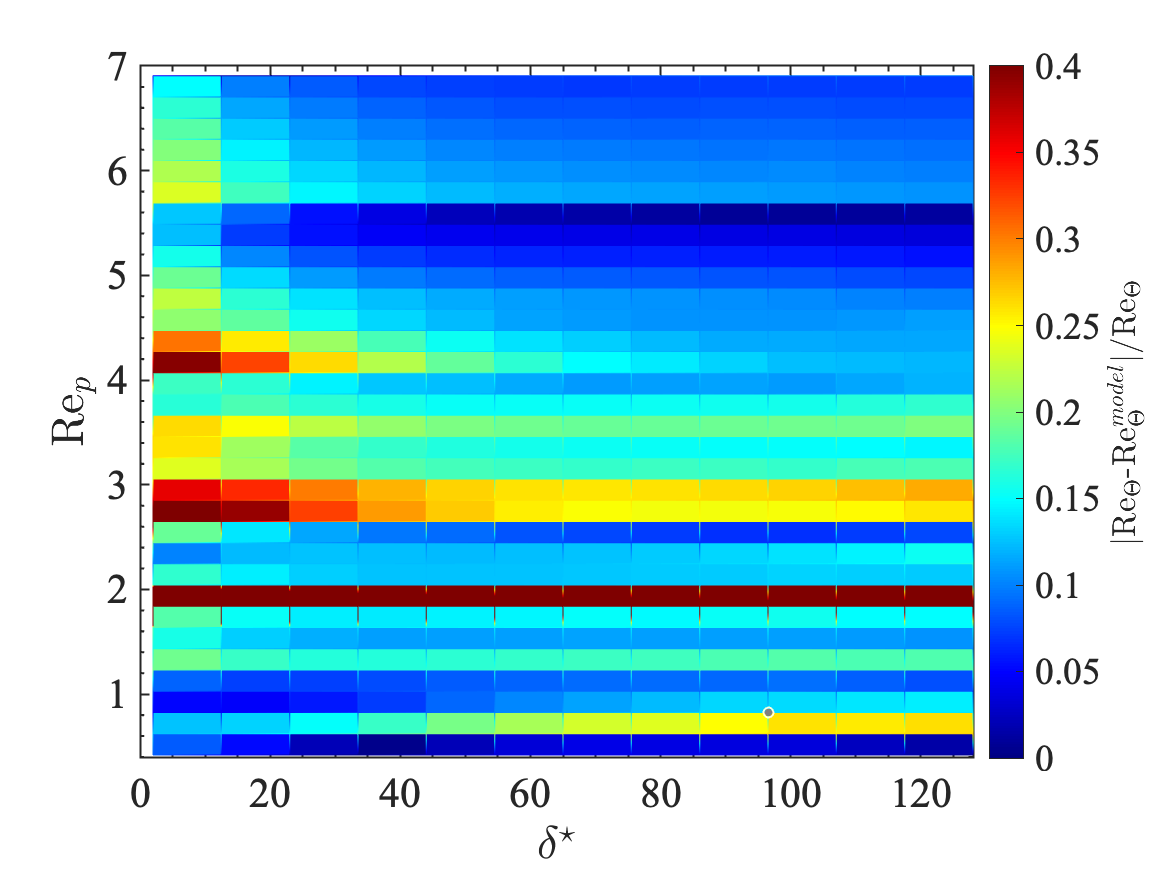}}
\caption{Comparison of the dependence of Re$_{\Theta}$ on Re$_p$ and $\delta^{\star}$ for CFD-DEM data (a), the model shown in Eq.~\ref{eq:powReg} (b), and the relative error. Note that the global mean error of the model is 20\% and the root mean squared error is 0.21.}
\label{fig:ReTcolormap}
\end{figure}

Because Re$_\Theta$ depends upon both Re$_p$ and $\delta^{\star}$, comparing the proposed model with the ground truth CFD-DEM data can be represented as either a 3D surface (see Fig.~\ref{fig:ReTSurface}) or a 2D color plot (see Fig.~\ref{fig:ReTcolormap}). In the former, we demonstrate the spread of the data both above and below the surface of the model prediction. It is notable that the computational campaign that serves as the basis for model development varied both volume fraction as well as Archimedes number. As previously discussed, the dependence of granular temperature on Archimedes number was weak, though this dependence introduces a degree of stochasticity. This can be seen both in the scatter of the CFD-DEM data in the 3D plot of Fig.~\ref{fig:ReTSurface} as well as in the 2D color plot of Fig.~\ref{fig:2Da}. 

We emphasize here that the original model of \citet{tang2016direct}, which was derived based on \emph{homogeneous} PR-DNS data, only needed to be recast to be representative of the \emph{heterogeneous} CIT data studied in this work and includes dependence upon filter size. However, it is important to note that the two model coefficients, $A_{\Theta},B_{\Theta}$, do not approach those of the original Tang model, $A_{\Theta},B_{\Theta} = (2.108, 0.85)$, when $\delta^{\star}\to0$, since a heterogeneous CIT state remains in the large-scale CFD-DEM data as the filter size approaches zero \cite{tang2016direct}.

The proposed model has a global mean error, defined as 
\begin{equation}
\mathcal{E}_{\text{gm}} = \frac{\vert\text{Re}_{\Theta}-\text{Re}_{\Theta}^{\text{model}} \vert}{\text{Re}_{\Theta}}
\end{equation} 
of 20\% relative to the underlying CFD-DEM data (the RMSE is 21\%). As shown in Fig.~\ref{fig:2Dc}, this is largely due to larger errors in regions where Archimedes number-based deviations occur in the data.

\section{Conclusion}
\label{sec:conclusions}
In this work, we present the most extensive, highly-resolved CFD-DEM study of granular temperature to date. This study considers the behavior of granular temperature with consideration given to volume fraction, Archimedes number and filtering size. As a part of this work, we propose a more widely applicable closer for the degree of clustering parameter, $\mathcal{D}$, which now can capture clustering behavior for gravity-driven gas-solid flows up to the limit of close packing. Further, we observe that the PR-DNS model for granular temperature, originally developed by \citet{tang2016direct}, correctly predicted the functional dependence between Re$_{\Theta}$ and Re$_p$. However, the coefficients informed by PR-DNS fall short of being predictive for large-scale behavior that captures mesoscale behavior such as clustering. As such, we propose \emph{filter-dependent} closures for the original coefficients in \citet{tang2016direct} and achieve marked improvement over the existing model. This work represents an initial step in developing improved closures to enable more accurate coarse-grained simulations of strongly-coupled, gas-solid flows, and future work is aimed at developing similarly filter-dependent closures for other important parameters such as particle drag. 

% wdf 5/8/25 - I should have included this in IJMF submission
\section*{Acknowledgments}
The authors would like to thank Jordan Musser (NETL) for developing the \texttt{postmfix} application used in this work and Manan Raval (MSIIP) for preliminary granular temperature modeling studies. This work was performed in support of the U.S. Department of Energy's (DOE) Office of Fossil Energy's Advanced Energy Systems Multi-Year Research Plan (MYRP) through the National Energy Technology Laboratory (NETL) Research \& Innovation Center's Simulation-Based Engineering (SBE) subprogram. This research used resources of the National Energy Research Scientific Computing Center, which is supported by the Office of Science of the U.S. Department of Energy under Contract No. DE-AC02-05CH11231.

\begin{appendices}
\section{Abbreviations and Symbols}
\label{appx:abbr}

%{\setlength{\extrarowheight}{0pt}

\begin{nomenclature}
\EntryHeading{Roman letters}
\entry{$\text{d}_{p}$}{Particle diameter}{[$\mu$m]}
\entry{$\bm{g}$}{Gravity}{[m s$^{-2}$]}
\entry{$\text{k}_{p}$}{Particle phase TKE}{[m$^{2}$ s$^{-2}$]}
\entry{$\text{k}_{f}$}{Fluid phase TKE}{[m$^{2}$ s$^{-2}$]}
\entry{$\text{m}_{p}$}{Particle mass}{[kg]}
\entry{$S$}{Entropy}{[m$^{2}$ s$^{-2}$ K$^{-1}$]}
\entry{$t$}{Time}{[s]}
\entry{$t_{0}$}{Time at onset of statistical steady state}{[s]}
\entry{$t_{f}$}{Final time of simulation}{[s]}
\entry{$U$}{Internal energy}{[m$^{2}$ s$^{-2}$]}
\entry{C$_{g}$}{Gravity coefficient}{[-]}
\entry{$\mathcal{D}$}{Degree of clustering}{[-]}
\entry{e$_{pp}$}{Restitution coefficient}{[-]}
\entry{$\mathcal{E}_{\text{gm}}$}{Global mean error}{[-]}
\entry{k$_{n}$}{Collisional spring constant}{[-]}
\entry{N$_{p}$}{Number of particles}{[-]}

\EntryHeading{Greek letters}
\entry{$\delta$}{Filter width }{[$\mu$m]}
\entry{$\Theta$}{Granular temperature }{[m$^{2}$ s$^{-2}$]}
\entry{$\kappa_{p}$}{Total granular energy }{[m$^{2}$ s$^{-2}$]}
\entry{$\mu_{g}$}{Gas Dynamic viscosity}{[Pa s]}
\entry{$\rho_{g}$}{Gas density}{[kg m$^{-3}$]}
\entry{$\rho_{p}$}{Particle density}{[kg m$^{-3}$]}
\entry{$\tau_p$}{Particle response time }{[s]}
\entry{$\varepsilon_{p}$}{Particle volume fraction }{[-]}
\entry{$\mu_{pp}$}{Interparticle friction coefficient }{[-]}
\entry{$\varphi$}{Mass loading }{[-]}

\EntryHeading{Dimensionless groups}
\entry{$Ar$}{Archimedes number }{${\rho_g \Delta \rho \left|{\bf g}\right| d_p^3}/{\mu_g^2}$}
\entry{Re}{Reynolds number}{$\rho_{g}\text{d}_{p}\bm{u} / \mu_{g}$}
\entry{$\text{Re}_{p}$}{Particle Reynolds number}{(See Eq. \ref{eq:rep})}
\entry{$\text{Re}_{\Theta}$}{Granular Temperature Reynolds number}{$\rho_{g}\text{d}_{p}\sqrt{\langle\Theta\rangle_{p}} / \mu_{g}$}
\EntryHeading{Miscellaneous dimensionless quantities}
\entry{$\delta^{\star}$}{Nondimensional filter width }{$\delta/\text{d}_{p}$}
\entry{$\delta^{\star}_{50}$}{Nondimensional filter width at which $3\langle\Theta_{p}\rangle/2=\kappa_{p}$ }{$\delta/\text{d}_{p}\ \vert\ 3\langle\Theta_{p}\rangle/2=\kappa_{p}$}
\entry{\tiny $\sqrt{\langle \varepsilon_p^{'2}\rangle}\vert_0$}{Volume fraction variance in uncorrelated particle assembly}{(see Eq. \ref{eq:clustDeg})}

\EntryHeading{Superscripts and subscripts}
\entry{${a}_{g}$}{Gas-specific quantity ``a''}{}
\entry{${a}_{p}$}{Particle-specific quantity ``a''}{}
\entry{${a}_{p}^{(i,t)}$}{Quantity ``a'' of i-th particle at time ``t''}{}

\EntryHeading{Miscellaneous notation}
\entry{$\widetilde{a}$}{Filtering operation on quantity $a$}{}
\entry{$\langle{a}\rangle$}{Reynolds average of quantity $a$}{$a=\langle a\rangle+a^{\prime}$}
\entry{$\langle{a}\rangle_{p}$}{Phase-weighted average of quantity $a$}{$\langle a \rangle_p = \frac{\langle \varepsilon_p a \rangle}{\langle \varepsilon_p\rangle}$}
\entry{${a}^{\prime}$}{Statistical fluctuations of quantity $a$}{$a^{\prime}=a -\langle{a}\rangle$}
\entry{$\tilde{{a}}_p^{''}$}{Statistical fluctuations about particle phase of quantity $a$}{$\tilde{{a}}_p^{''} = \tilde{{a}}_p - \langle \tilde{{a}}_p\rangle_p$}

\end{nomenclature}

\end{appendices}

\newpage

\bibliographystyle{plainnat} 
\bibliography{cas-refs}

\end{document}

